\documentclass[12pt]{article}
\usepackage{amsmath,amssymb,amsthm,amsxtra,overpic,bbm,bm,epsfig,subfigure}
\usepackage{hyperref}
\usepackage{mathrsfs}
\usepackage{graphicx}
\usepackage{multirow}
\usepackage{color}
\usepackage{comment}
\usepackage{epstopdf}
\numberwithin{equation}{section}
\usepackage{float}
\usepackage{cite}
\usepackage{hyperref}
\usepackage{url}
\usepackage{slashed,stmaryrd}
\usepackage{longtable}
\usepackage{extarrows}

\begin{document}

\def\thefootnote{\fnsymbol{footnote}}
\vspace{0.2cm}
\begin{center}
{\Large\bf Invisible Neutrino Decays as Origin of TeV Gamma Rays from GRB221009A}
\end{center}
\vspace{0.2cm}
	
\begin{center}
{\bf Jihong Huang}~\footnote{E-mail: huangjh@ihep.ac.cn},\quad
{\bf Yilin Wang}~\footnote{E-mail: wangyilin@ihep.ac.cn},\quad
{\bf Bingrong Yu}~\footnote{E-mail: yubr@ihep.ac.cn},\quad
{\bf Shun Zhou}~\footnote{E-mail: zhoush@ihep.ac.cn}
\\
\vspace{0.2cm}
{\small
Institute of High Energy Physics, Chinese Academy of Sciences, Beijing 100049, China\\
School of Physical Sciences, University of Chinese Academy of Sciences, Beijing 100049, China}
\end{center}
	
\vspace{0.5cm}
	
\begin{abstract}
Recently, the LHAASO collaboration has observed the gamma rays of energies up to ten TeV from the gamma-ray burst GRB221009A, which has stimulated the community of astronomy, particle physics and astrophysics to propose various possible interpretations. In this paper, we put forward a viable scenario that neutrinos are produced together with TeV photons in the gamma-ray burst and gradually decay into the axion-like particles, which are then converted into gamma rays in the galactic magnetic fields. In such a scenario, the tension between previous axion-like particle interpretations and the existing observational constraints on the relevant coupling constant and mass can be relaxed. 
\end{abstract}
	
\newpage
	
\def\thefootnote{\arabic{footnote}}
\setcounter{footnote}{0}

\section{Introduction}
\label{sec:intro}
A highly-energetic outburst has been detected by Swift-XRT~\cite{GCN32635}, and immediately confirmed as the gamma-ray burst GRB221009A by Fermi-GBM~\cite{GCN32636,GCN32642} and Fermi-LAT~\cite{GCN32637,GCN32658} within a duration of $T_{90}^{} = 327~{\rm s}$ and an event fluence $S \approx 18.2~{\rm GeV}~{\rm cm}^{-2}$ in the energy range $(10\cdots 1000)~{\rm keV}$. The redshift of GRB221009A is determined to be $z_0^{} \approx 0.15$ by X-shooter/VLT~\cite{GCN32648} and the GTC telescope~\cite{GCN32686}, corresponding to a comoving distance of $\ell\approx 643~{\rm Mpc}$ to the Earth. It is the brightest GRB ever detected. In particular, within 2000~{\rm s} after the burst, more than 5000 very-high-energy (VHE) photons above 500~GeV have been observed by the detectors of Large High Altitude Air Shower Observatory (LHAASO), where the photon of the highest energy reaching 18~TeV has been recorded by the KM2A detector~\cite{GCN32677}. In this connection, the Carpet-2 experiment has even claimed the detection of 251~TeV photon-like shower events after the trigger~\cite{Carpet2}. 


The observation of photons with energies above 10~TeV is striking, because the flux of such VHE photons would have been severely attenuated in the extragalactic background light (EBL) via the electron-positron pair production $\gamma + \gamma^{}_{\rm EBL} \to e^+_{} + e^-_{}$ and they could hardly arrive in the Milky Way (MW)~\cite{Gould:1966pza}. 
Although it is still possible to explain the ${\mathcal O}(10)~{\rm TeV}$ photons based on the standard propagation of VHE photons in the EBL~\cite{Zhao:2022wjg}, one has to properly choose the EBL model, which currently has a large uncertainty (see, e.g., Refs.~\cite{Baktash:2022gnf, Carenza:2022kjt} and references therein). To have a feel about the attenuation,  we take an average value of the optical depth $\tau^{}_{\rm op} \approx 15$ for $E^{}_\gamma = 18~{\rm TeV}$ for illustration and find accordingly the survival probability of ${\rm exp}(-15)\approx 10^{-7}_{}$.
Thus far, LHAASO's detection of the 18~TeV photon has triggered various new-physics interpretations,\footnote{However, it is worthwhile to mention that the energy resolution of LHAASO is about 40\% at 18~TeV~\cite{Cui:2014bda}, implying the lower bound of the energy around 10~TeV. If the real energy of the detected photon is close to 10~TeV, there will be no need to invoke any new-physics scenarios.} including the Lorentz invariance violation~\cite{Li:2022vgq,Baktash:2022gnf,Li:2022wxc,Finke:2022swf,Zhu:2022usw,He:2022jdl,Huang:2022xto,Vardanyan:2022ujc}, dark photon~\cite{Gonzalez:2022opy}, axion-like particles (ALPs)~\cite{Galanti:2022pbg,Baktash:2022gnf,Lin:2022ocj,Troitsky:2022xso,Nakagawa:2022wwm,Zhang:2022zbm,Gonzalez:2022opy,Carenza:2022kjt,Galanti:2022xok} and sterile neutrinos~\cite{Cheung:2022luv,Smirnov:2022suv,Brdar:2022rhc}. In addition, the interaction of ultra-high-energy cosmic rays with the cosmological radiation background may give a viable explanation~\cite{AlvesBatista:2022kpg,Das:2022gon}.

In this paper, we propose a viable scenario of invisible neutrino decays $\nu^{}_i \to \nu^{}_j + a$, where $\nu^{}_i$ and $\nu^{}_j$ are neutrino mass eigenstates (with masses $m^{}_i > m^{}_j$) and $a$ stands for an ALP, to explain LHAASO's observation of VHE gamma rays. First of all, the hadronic origin of TeV gamma rays from the GRB will be assumed such that TeV neutrinos are produced together with photons. As for GRB221009A, the follow-up observation of track-like events induced by muon neutrinos at IceCube has not found any signals~\cite{GCN32665,Abbasi:2023xhh}. Assuming an $E^{-2}_\nu$ spectrum in the energy range $800~{\rm GeV} < E^{}_\nu < 1~{\rm PeV}$, the IceCube observation leads to an upper bound on the time-integrated neutrino flux $E^2_\nu \Phi^{}_\nu < 3.9 \times 10^{-2}~{\rm GeV}~{\rm cm}^{-2}$ at the 90\% confidence level. This upper bound should be taken into account. Second, as the high-energy neutrinos come out of the GRB and propagate to the Earth, they will gradually decay into the lightest neutrinos and ALPs. Finally, these energetic ALPs convert to VHE photons via a two-photon interaction vertex in the magnetic fields of the MW, which are detected by LHAASO. As will be shown later, even under very restrictive bounds on the lifetime of neutrinos from cosmology~\cite{Barenboim:2020vrr}, the rate of neutrino decays into ALPs is sizable enough. Remarkably, such a scenario can naturally predict the observed VHE photons around 18~TeV with an event number of ${\cal O}(1)$ at LHAASO, while satisfying the IceCube constraint~\cite{GCN32665,Abbasi:2023xhh}. In comparison with the previous works on the ALP interpretation~\cite{Galanti:2022pbg,Baktash:2022gnf,Lin:2022ocj,Troitsky:2022xso,Nakagawa:2022wwm,Zhang:2022zbm,Gonzalez:2022opy,Carenza:2022kjt,Galanti:2022xok}, the desired ALP-photon coupling constant in our scenario turns out to be well consistent with all the existing observational limits.

The remaining part of this paper is organized as follows. In Sec.~\ref{sec:flux}, we compute the fluxes of high-energy neutrinos and gamma rays from GRB221009A by adopting a conventional fireball model and a simplified treatment of photohadronic interactions. The decay rate of $\nu^{}_i \to \nu^{}_j + a$ over the cosmological distance and the conversion probability from ALPs to photons in the external magnetic field are calculated in Sec.~\ref{sec:invisible_decay} and Sec.~\ref{sec:ALP-photon_converison}, respectively. The expected number of photons that can be detected by LHAASO is calculated in Sec.~\ref{sec:photon_numbers}. Finally, our main results are summarized in Sec.~\ref{sec:summary}.

\section{TeV Neutrinos and Photons from GRBs}
\label{sec:flux}
The GRBs are expected to be the sources of high-energy cosmic rays~\cite{Piran:1999kx, Kumar:2014upa}, and the accelerated protons will interact with ambient photons, producing TeV neutrinos and gamma rays via photohadronic processes and subsequent decays~\cite{Waxman:1997ti,Waxman:1998yy}. In the conventional fireball model of GRBs~\cite{Piran:1999kx}, the central engine emits matter shells of thickness $\Delta r \approx c \, t^{}_{\rm v}/(1+z)$, where $c$ is the speed of light, $z$ the redshift and $t^{}_{\rm v}$ the time variability in the GRB observation in the observer's frame, and the thermal photons from the fireball transfer their energies to the baryons in the shell. The ultimate Lorentz factor of the accelerated shell can be estimated as $\Gamma = {\cal E}^{}_{\rm tot}/\left({\cal M} c^2\right)$ with ${\cal E}^{}_{\rm tot}$ being the total energy of the fireball and ${\cal M}$ being the total mass of baryons. Different shells start to collide with each other at the radius around $r^{}_{\rm C} \approx 2 \Gamma^2 c \, t^{}_{\rm v}/(1+z)$, where these internal collisions lead to an efficient acceleration of protons. 

Closely following Refs.~\cite{Hummer:2010vx,Baerwald:2013pu}, we now calculate the fluxes of high-energy neutrinos and photons from GRBs by adopting the simplified treatment of photohadronic interactions. 
\begin{itemize}
\item First, the energy spectrum of background photons is taken to be of a broken power law: $n^\prime_\gamma = C^{}_\gamma (E^\prime_\gamma/E^\prime_{\gamma, {\rm b}})^{-\alpha^{}_\gamma}$ for $E^\prime_{\gamma, {\rm min}} \leqslant E^\prime_\gamma < E^\prime_{\gamma, {\rm b}}$, and $n^\prime_\gamma = C^{}_\gamma (E^\prime_\gamma/E^\prime_{\gamma, {\rm b}})^{-\beta^{}_\gamma}$ for $E^\prime_{\gamma, {\rm b}} \leqslant E^\prime_\gamma < E^\prime_{\gamma, {\rm max}}$ and $n^\prime_\gamma = 0$ otherwise, where all the primed quantities refer to those in the shock rest frame (SRF), and $n^\prime_\gamma$ is given in units of ${\rm GeV}^{-1}~{\rm cm}^{-3}$. Furthermore, the power indices $\alpha^{}_\gamma = 1$, $\beta^{}_\gamma = 2$ and the break energy $E^\prime_{\gamma, {\rm b}} = (1~{\rm MeV})/\Gamma$ will be taken as usual. Here we fix the break energy of target photons as $E_{\gamma,{\rm b}} = 1~{\rm MeV}$ in the observer's frame, and a Lorentz factor $\Gamma$ should be divided when changing to the SRF. The minimal and maximal photon energies are set to $E^\prime_{\gamma, {\rm min}} = 0.2~{\rm eV}$ and $E^\prime_{\gamma, {\rm max}} = 300~{\rm keV}$, respectively. The normalization constant $C^{}_\gamma$ can be determined from the observed gamma-ray fluence, which is $S \approx 18.2~{\rm GeV}~{\rm cm}^{-2}$ in the energy range $10~{\rm keV} \lesssim E^{}_\gamma \lesssim 1~{\rm MeV}$ for GRB221009A. More explicitly, we have~\cite{Baerwald:2013pu}
\begin{eqnarray}
    \int^{(1~{\rm MeV})/\Gamma}_{(10~{\rm keV})/\Gamma} E^\prime_\gamma n^\prime_\gamma\ {\rm d}E^\prime_\gamma = \frac{E^{\prime {\rm sh}}_{\rm iso}}{V^\prime_{\rm iso}} \; ,
    \label{eq:gammanorm}
\end{eqnarray}
where $V^\prime_{\rm iso} \approx 4\pi r^2_{\rm C} (\Gamma \Delta r)$ denotes the isotropic volume for particle production in the shell and $E^{\prime {\rm sh}}_{\rm iso} = E^{\rm sh}_{\rm iso}/\Gamma$ with $E^{\rm sh}_{\rm iso} \approx 4\pi d^2_{\rm L} (F^{}_\gamma t^{}_{\rm v})/(1+z)$ being the isotropic equivalent energy per shell. Note that here $d^{}_{\rm L}$ is the luminosity distance and $F^{}_\gamma = S/T^{}_{90}$ is the radiative flux. Given the redshift $z^{}_0 \approx 0.15$ for GRB221009A, the luminosity distance is found to be $d^{}_{\rm L} \approx 739~{\rm Mpc}$ with the input of standard cosmological parameters. For illustration, we shall take $t^{}_{\rm v} = 0.1~{\rm s}$ and different values of the Lorentz factor $\Gamma$ in the following discussions.

\item Second, we assume that the energy spectrum of accelerated protons from internal collisions in the SRF follows a power law with a cutoff: $n_p^\prime = C_p^{} \left(E_p^\prime/E_{p,{\rm min}}^\prime \right)^{-\alpha_p^{}}_{} {\rm exp} \left[-\left(E_p^\prime/E_{p,{\rm max}}^\prime \right)^k_{} \right] $ for $E_{p,{\rm min}}^\prime \leqslant E_p^\prime$ and $n_p^\prime = 0$ otherwise, where $n_p^\prime$ is given in units of ${\rm GeV}^{-1}~{\rm cm}^{-3}$. The minimal proton energy in the SRF is $E_{p,{\rm min}}^\prime = 1 ~ {\rm GeV}$, whereas the maximal energy $E_{p,{\rm max}}^\prime$ depends on the acceleration efficiency, the cooling rate and the photohadronic interaction rate. Here $E_{p,{\rm max}}^\prime = 6.9 \times 10^8 ~ {\rm GeV}$ is fixed as a benchmark value for simplicity as in Refs.~\cite{Baerwald:2013pu,Lipari:2007su}. In addition, the power index $\alpha_p^{}=2$ and $k = 1$ in the exponential cutoff are chosen. Introducing the ratio $f^{}_e$ between the energy of electrons and that of protons, one can extract the normalization coefficient $C^{}_p$ from the observed gamma-ray fluence by further assuming the energy equipartition between electrons and photons~\cite{Baerwald:2013pu}, namely, 
\begin{eqnarray}
\int_{E_{p,{\rm min}}^\prime}^{E_{p,{\rm max}}^\prime} E_p^\prime \, n_p^\prime \ {\rm d}  E_p^\prime = \frac{1}{f_e^{}} \frac{E_{\rm iso}^{\prime {\rm sh}}}{V_{\rm iso}^{\prime}} \;.
\label{eq:protonnorm}
\end{eqnarray}
The calculation of $E^{\prime {\rm sh}}_{\rm iso}$ and $V^\prime_{\rm iso}$ has been explained below Eq.~(\ref{eq:gammanorm}). The baryon loading parameter $f^{-1}_e$ affects the yields of secondary particles in the photohadronic interactions in a significant way, so it will be allowed to vary in the wide range of $f^{-1}_e \in [5, 100]$ or equivalently $f^{}_e \in [0.01, 0.2]$~\cite{Hummer:2011ms}.

\item Finally, given the photon spectrum $n^\prime_\gamma$ and the proton spectrum $n^\prime_p$ in the SRF, one can compute the injection spectra $Q^\prime(E^\prime)$ of any secondary particle from the photohadronic interactions and subsequent decays. In our calculations, we implement the approximate but reasonable modeling of photohadronic interactions first given in Ref.~\cite{Hummer:2010vx}, where the cross sections of the $p$-$\gamma$ interactions in the dominant channels, including the $\Delta$-resonance, the higher resonances of nucleon states, direct production of pions and multi-pion production, and the decay kinematics are carefully modeled and tabulated. Making use of the ``Sim-B" model of photohadronic interactions in Ref.~\cite{Hummer:2010vx}, we calculate the injection spectra of pions, kaons and neutrons. The subsequent decays of charged pions and kaons, and neutrons give rise to high-energy neutrinos, while neutral pions decay into high-energy gamma rays. As a result, we obtain the injection spectra of high-energy neutrinos $Q^\prime_{\nu^{}_e}(E^\prime_\nu)$ and $Q^\prime_{\nu^{}_\mu}(E^\prime_\nu)$, and that of photons $Q^\prime_\gamma(E^\prime_\gamma)$, where neutrinos and antineutrinos of each flavor will not be distinguished in this work. Then, the injection spectrum $Q^\prime(E^\prime)$ is utilized to derive the flux via $\phi(E^\prime) = V_{\rm iso}^\prime Q^\prime (1+z)^2/(4\pi d_{\rm L}^2)$, which will be converted to the flux $\phi(E)$ in the observer's frame via $E = \Gamma E^\prime/ (1+z)$. It is worth mentioning that the shell collisions at different radii are assumed to be similar for particle production and the number of shells can be estimated as $N \approx T_{90}^{} / t_{\rm v}^{}$. The contributions from all the shells are equally taken into account. 
\end{itemize}
\begin{figure}[t!]
	\centering
	\includegraphics[scale=0.62]{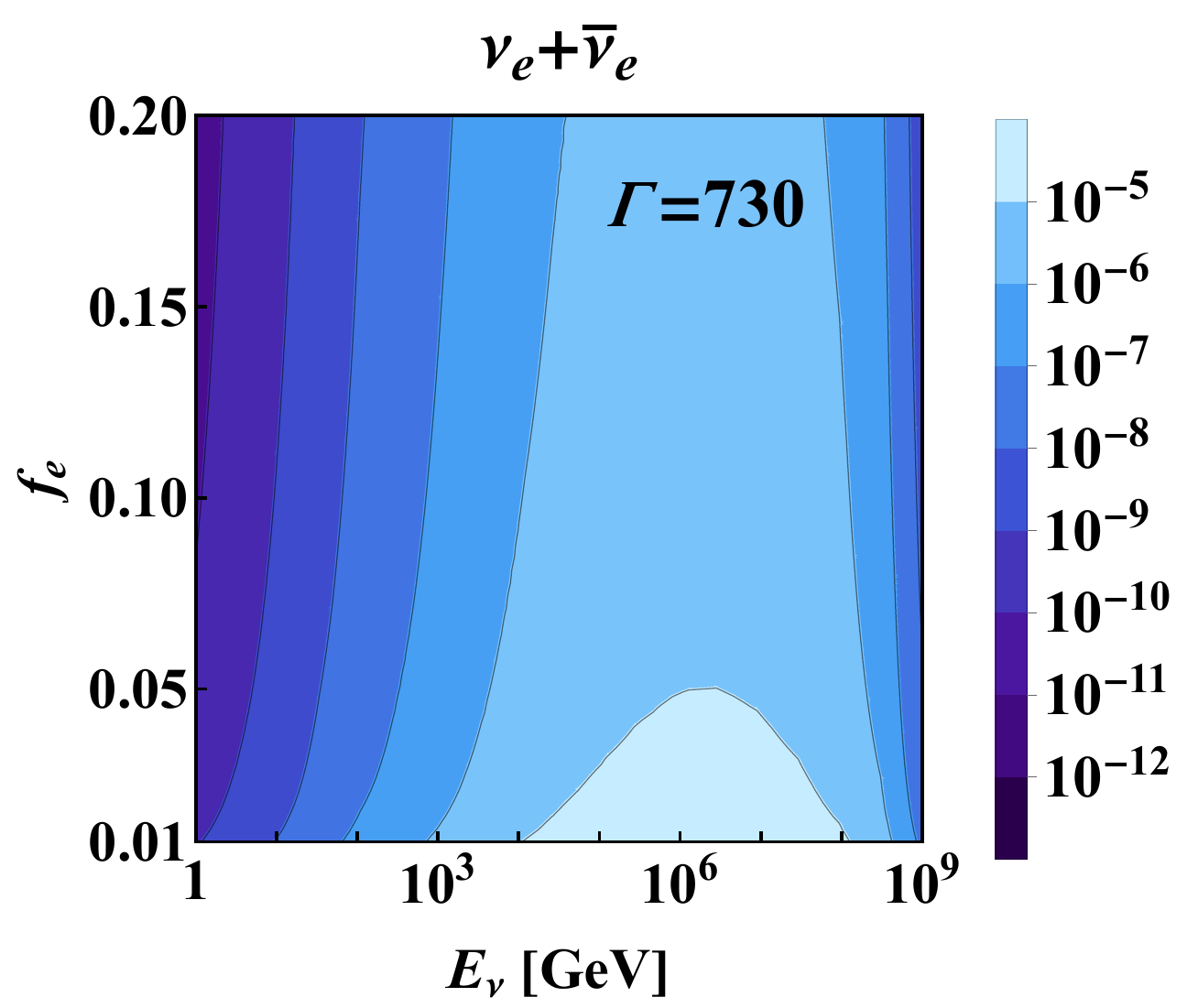}\quad
	\includegraphics[scale=0.62]{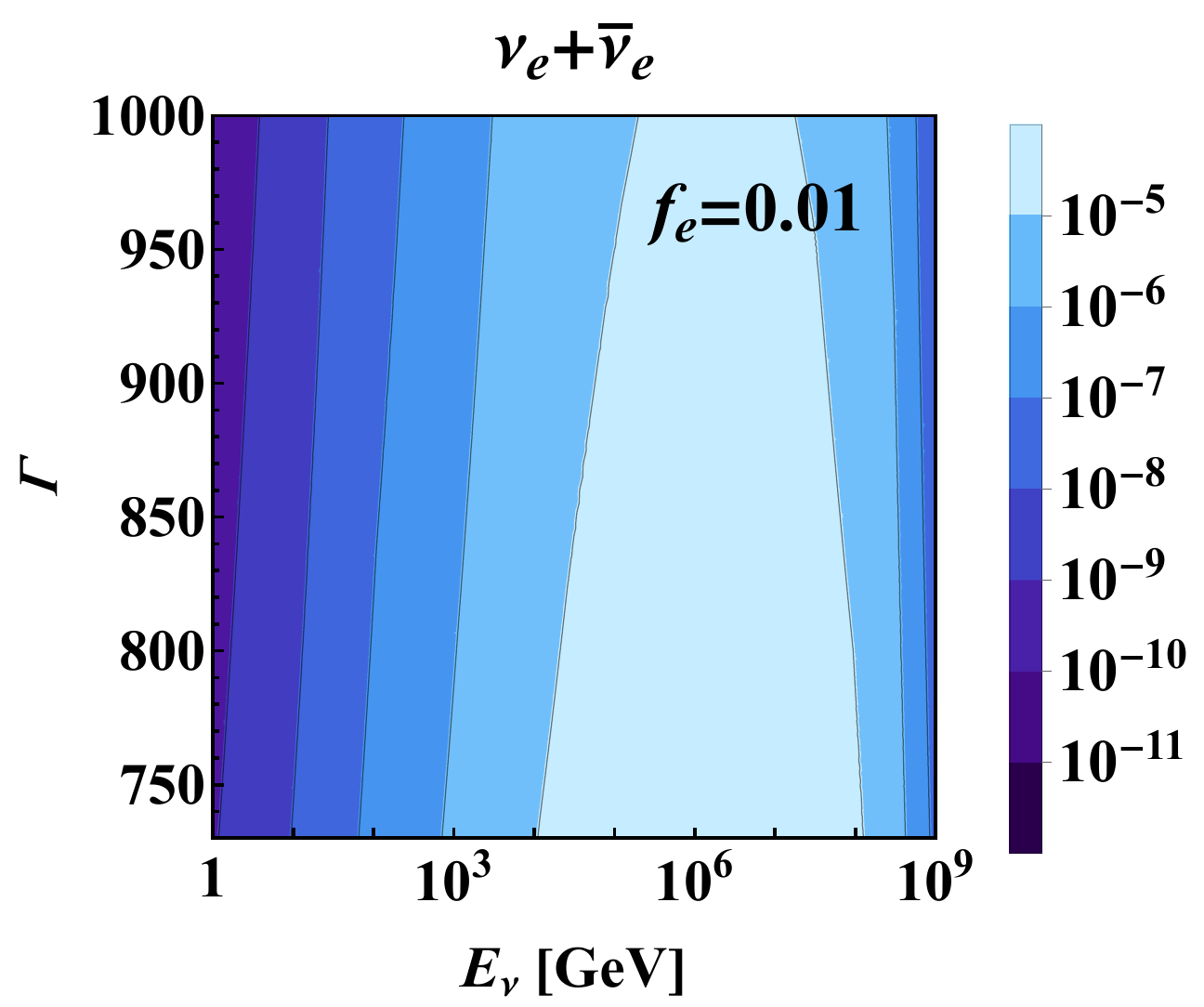}\\
        \vspace{0.15cm}
        \includegraphics[scale=0.62]{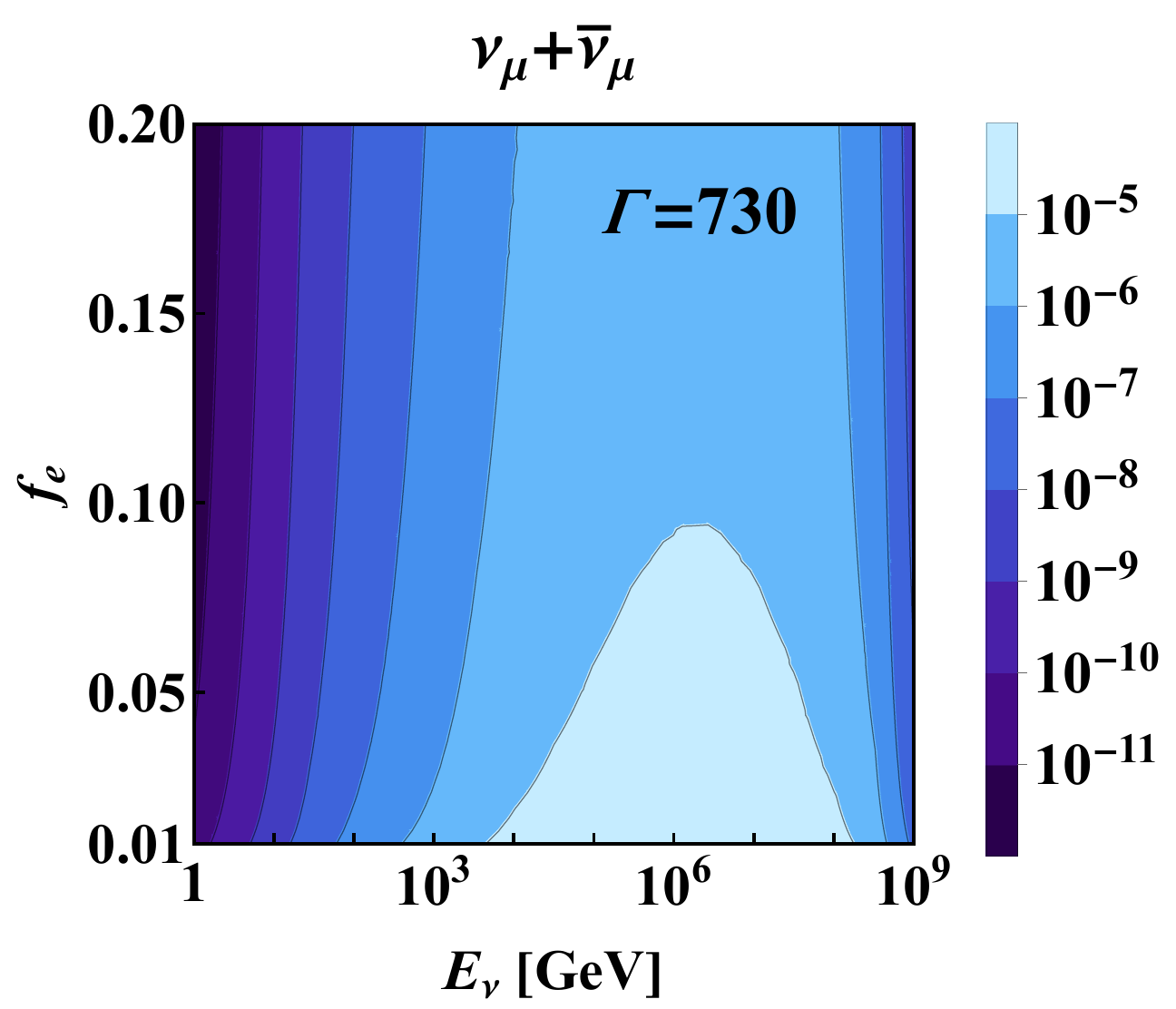}\quad
	\includegraphics[scale=0.62]{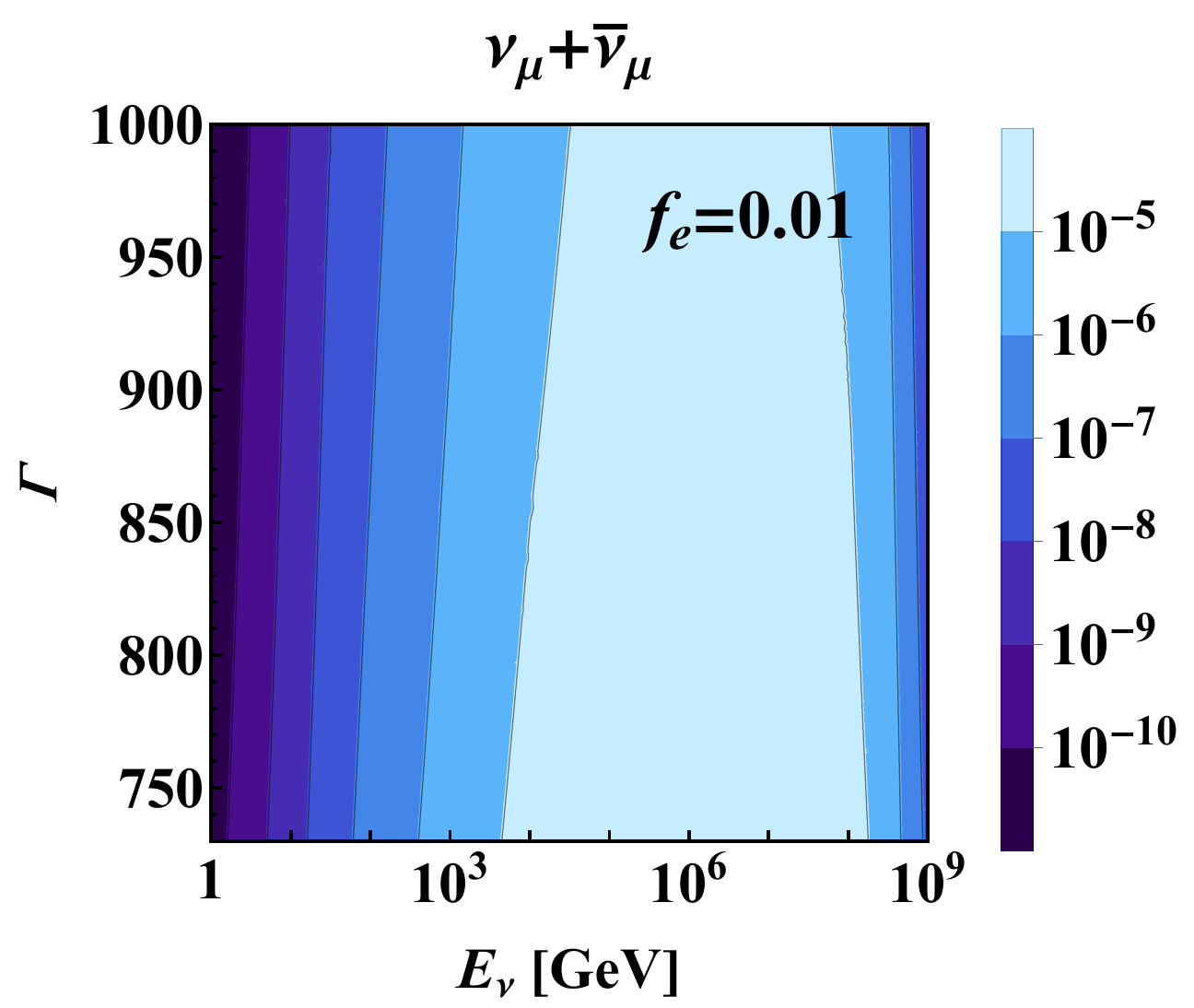}\\
        \vspace{0.05cm}\,
         \includegraphics[scale=0.62]{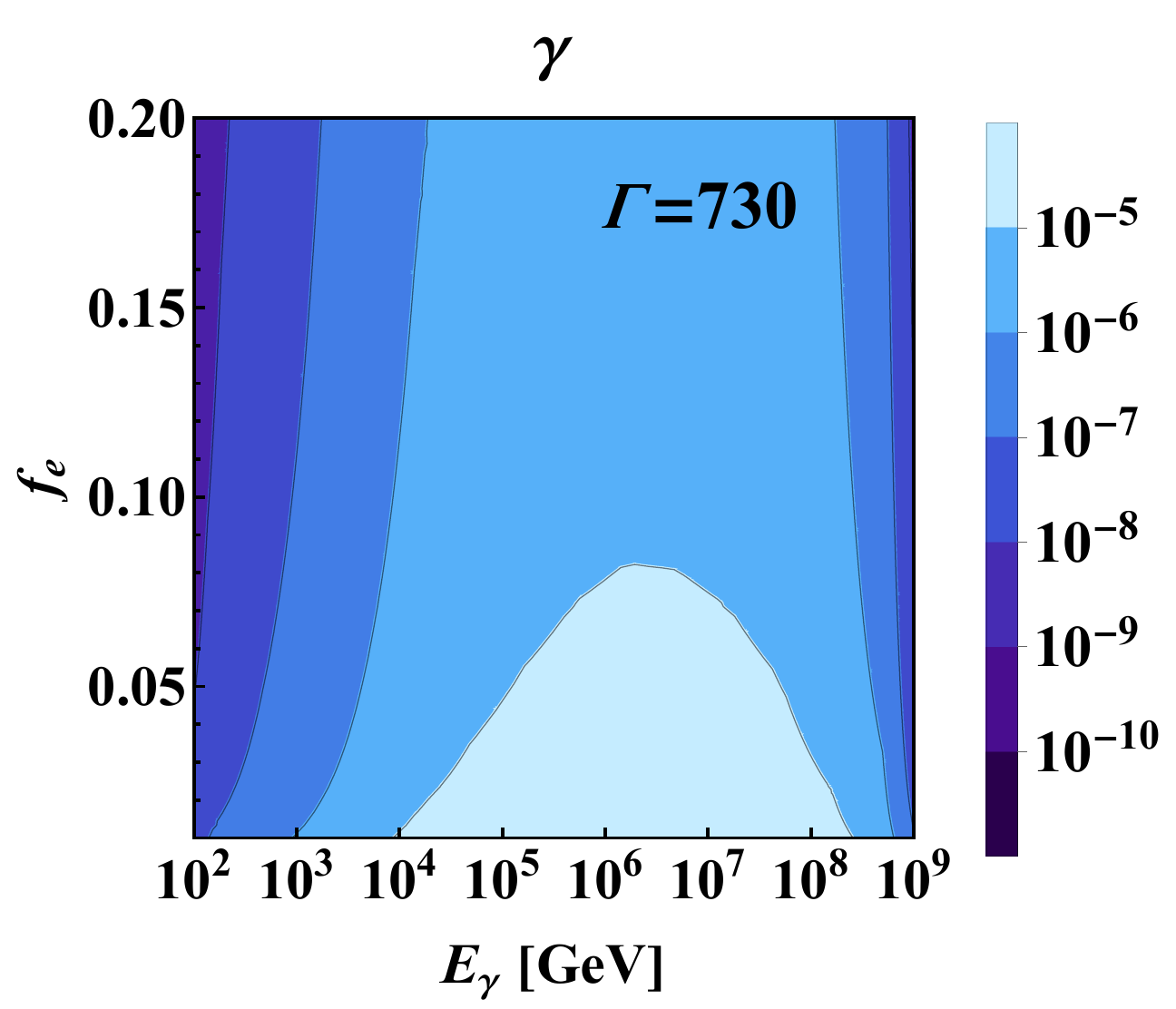}\quad
	\includegraphics[scale=0.62]{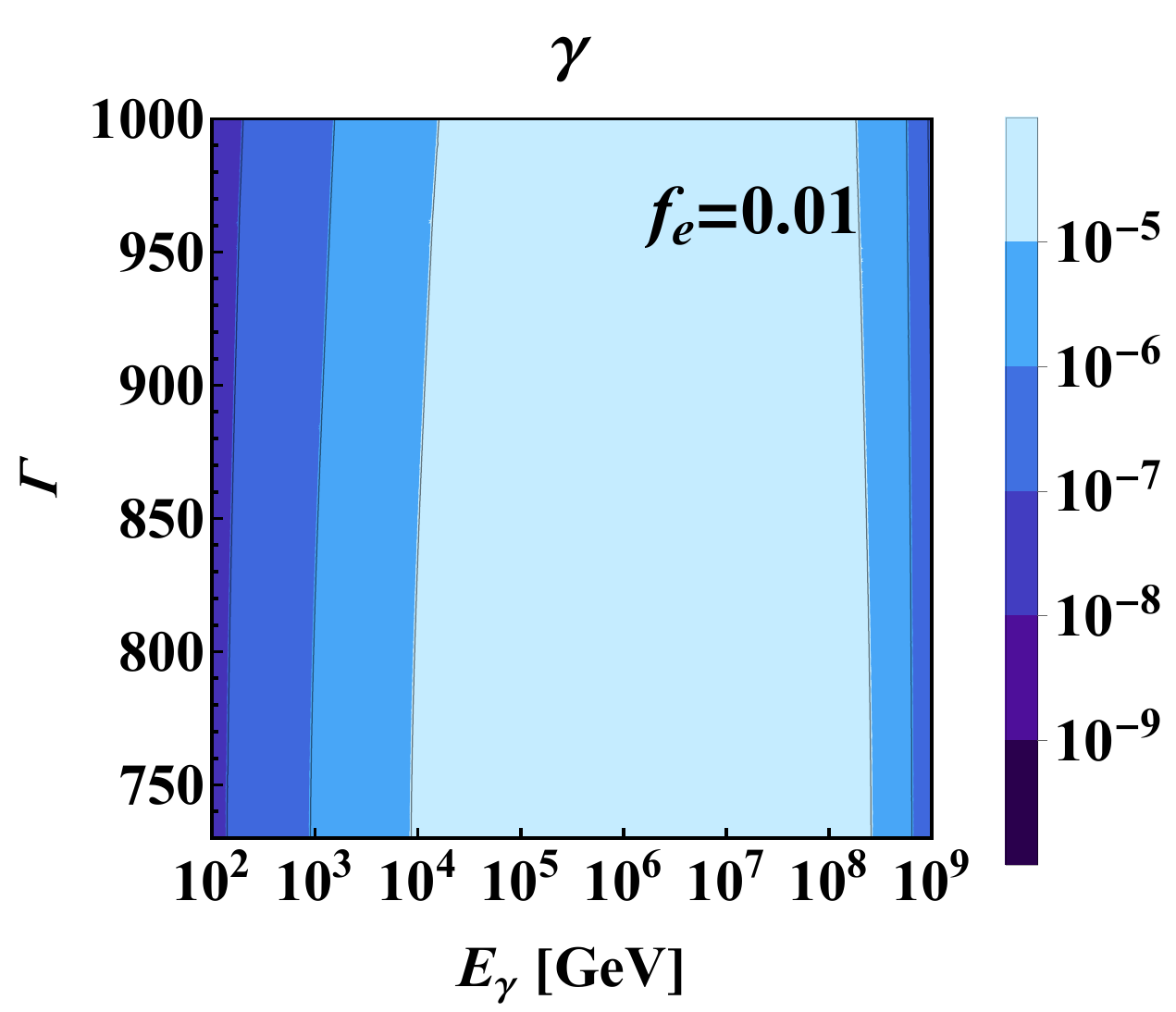}
	\caption{The fluxes (in units of ${\rm GeV}~{\rm cm}_{}^{-2}~{\rm s}_{}^{-1}$) of electron neutrinos (top row), muon neutrinos (middle row) and photons (bottom row) for various values of the electron-to-proton energy ratio $f_e^{}$ (left column) and the Lorentz factor $\Gamma$ (right column), where the fluxes of neutrinos and antineutrinos are added up. The time variability $t_{\rm v}^{} = 0.1~{\rm s}$ in the GRB observation is fixed.}
	\label{fig:contour}
\end{figure}
\begin{figure}[t!]
	\centering
	\includegraphics[scale=0.7]{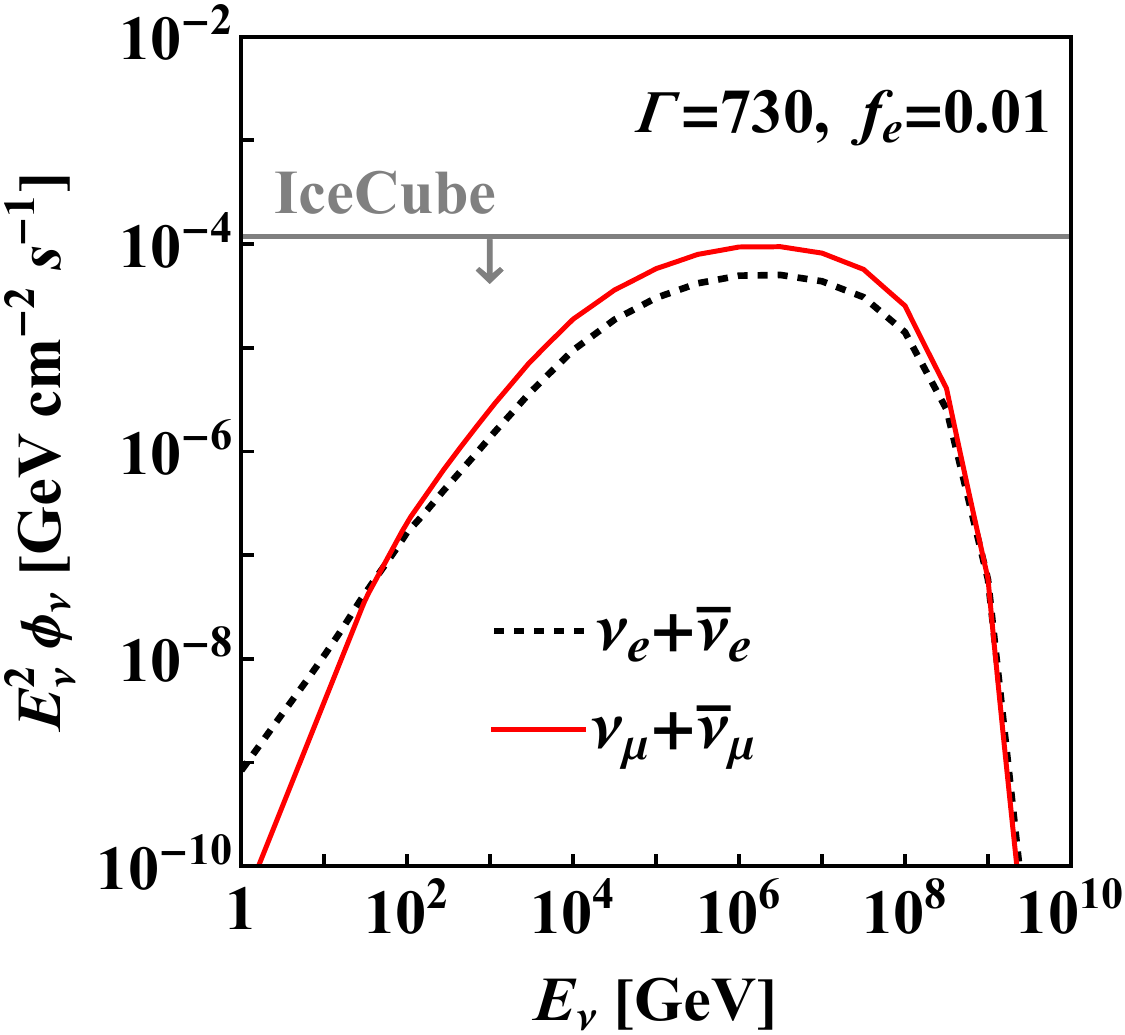}\quad
    \includegraphics[scale=0.7]{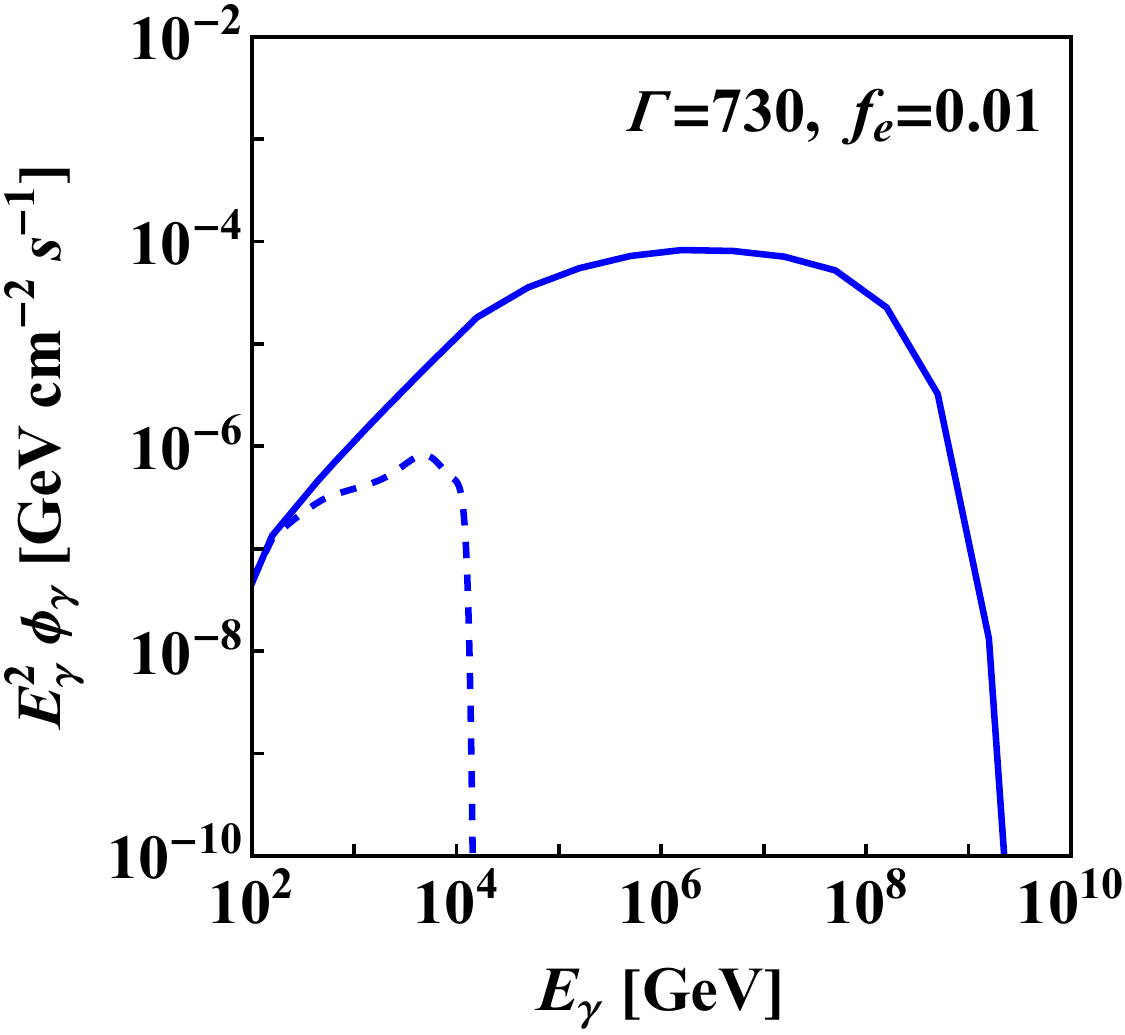}
	\caption{The fluxes of neutrinos (left panel) and photons (right panel) with the Lorentz factor $\Gamma =730$ and the electron-to-proton energy ratio $f_e^{} = 0.01$. Electron neutrinos and muon neutrinos are shown as black dotted curve and red solid curve, respectively, where the fluxes of neutrinos and antineutrinos have been added up. The horizontal gray line shows the upper bound derived from the non-observation of muon-track events in IceCube, which has been averaged by $T_{90}^{}=327~{\rm s}$~\cite{GCN32665}.  Notice that the neutrino fluxes in the left panel are calculated at the source, whereas the photon fluxes in the right panel are calculated both at the source (solid curve) and at the detector (dashed curve) by using the EBL model from Ref.~\cite{Franceschini:2008tp}.}
	\label{fig:Enu}
\end{figure}

Notice that the high-energy photons will be attenuated, when propagating out of the GRB, by the background photons via the electron-positron pair production $\gamma+\gamma\to e_{}^{+}+e_{}^{-}$. Such an attenuation effect in the internal shock should also be taken into account when calculating the flux of high-energy photons. For smaller values of the Lorentz factor $\Gamma$, the energies of background photons in the internal shock will be higher such that the rate of pair production is larger. This observation leads to a lower bound on the Lorentz factor $\Gamma \gtrsim 730 $~\cite{Murase:2022vqf}, for which the optical depth of high-energy photons in the emission region becomes smaller than 1. For this reason, we impose such a lower bound on the Lorentz factor and take account of the attenuation of high-energy photons from $\pi^0$ decays in our calculations.

Following the above strategy, we obtain the fluxes of electron and muon neutrinos, as well as high-energy photons, from a specific model of GRB221009A. The final numerical results are shown in Fig.~\ref{fig:contour}, where the Lorentz factor $\Gamma$ and the electron-to-proton energy ratio $f_e^{}$ are allowed to vary in the range of $\Gamma \in [730, 1000]$ and $f_e^{} \in [0.01, 0.2]$. Some comments on the numerical results are in order. One can observe from Eqs.~(\ref{eq:gammanorm}) and (\ref{eq:protonnorm}) that the energy densities of photons and protons are inversely proportional to $\Gamma^6$ and thus decrease rapidly for an increasing value of the Lorentz factor. Consequently, the fluxes of high-energy neutrinos and photons become smaller for larger values of $\Gamma$. 
In the meanwhile, according to Eq.~(\ref{eq:protonnorm}), the normalization coefficient $C_p^{}$ which appears in the energy densities is inversely proportional to $f_e^{}$, thus an increasing value of $f_e^{}$ also reduces the fluxes. The fluxes of neutrinos and photons with the fixed value of $\Gamma=730$ and $f_e^{}=0.01$ are shown in Fig.~\ref{fig:Enu}, which can be regared as the maximum fluxes in the allowed range of $\Gamma$ and $f_e^{}$. In the left panel of Fig.~\ref{fig:Enu}, the upper bound derived from the null signal of muon-track events coincident with GRB221009A at IceCube~\cite{GCN32665} is displayed as the horizontal line. One can immediately see that $\Gamma=730$ and $f_e^{} = 0.01$ lead to the muon neutrino fluxes compatible with the IceCube bound. According to this lower bound on the Lorentz factor $\Gamma \gtrsim 730$, we can further estimate the typical radius of internal shocks $r^{}_{\rm C} \gtrsim 2.8\times 10^{15}~{\rm cm}$, which is consistent with the results in Ref.~\cite{Ai:2022kvd,Murase:2022vqf,Liu:2022mqe}. 

\section{Invisible Neutrino Decays}
\label{sec:invisible_decay}
When the standard model (SM) is extended with nonzero neutrino masses, heavier neutrinos could decay into lighter ones and photons $\nu^{}_i \to \nu^{}_j + \gamma$ at the one-loop level, where $\nu^{}_i$ and $\nu^{}_j$ are neutrino mass eigenstates with masses $m^{}_i > m^{}_j$. However, the rates of such radiative neutrino decays are highly suppressed for the SM interactions and neutrino lifetimes are thus much longer than the age of our Universe, satisfying the experimental constraints from direct searches for solar gamma rays~\cite{Raffelt:1985rj} and from the spectral distortions of the cosmic microwave background (CMB)~\cite{Aalberts:2018obr}. In connection with neutrino mass generation, new physics beyond the SM is indispensable. 

Instead of specifying any concrete new-physics models~\cite{Chikashige:1980ui,Gelmini:1980re,Georgi:1981pg,Dvali:2013cpa,Dvali:2016uhn,Zhou:2007zq,Huang:2018cwo}, we shall consider the following effective interactions between neutrinos and a pseudo-scalar particle, i.e.,
\begin{eqnarray}
\label{eq:Lagint}
{\cal L}_{\nu}=\sum_{\alpha,\beta}^{}{\rm i}\tilde{y}_{\alpha \beta}^{}\overline{\nu_\alpha^{}}\gamma_5^{}\nu_\beta^{}a+{\rm h.c.}=\sum_{i,j}^{}{\rm i }y_{ij}\overline{\nu_i^{}}\gamma_5^{}\nu_j^{}a+{\rm h.c.}\;,
\end{eqnarray}
where $a$ is the pseudo-scalar of mass $m_a^{}$, $\tilde{y}_{\alpha\beta}^{}$ with $\alpha,\beta=e,\mu,\tau$ are the couplings in the flavor basis, while $y_{ij}^{}=\sum_{\alpha,\beta}^{}U_{\alpha i}^{*}U_{\beta j}^{}\tilde{y}_{\alpha \beta}^{}$ are those in the mass basis. Here $U$ denotes the unitary flavor mixing matrix in the leptonic sector. Notice that the pseudo-scalar particle $a$ will be regarded as an ALP, which additionally interacts with photons as below
\begin{eqnarray}
 {\cal L}^{}_a = \frac{1}{2} \left(\partial_\mu a \partial^\mu a - m_a^2 a^2_{}\right) + \frac{g_{a\gamma}^{}}{4} F_{\mu\nu}^{} \tilde{F}^{\mu\nu}_{} a \; ,
 \label{eq:Laga} 
\end{eqnarray}
where $\tilde{F}^{\mu\nu}$ stands for the dual of the electromagnetic field strength tensor $F^{\mu\nu}$.
For the moment, we assume the ALP-photon interaction characterized by the coupling $g^{}_{a\gamma}$ to be weak enough such that the decays $\nu^{}_i \to \nu^{}_j + a$ (for $m^{}_i > m^{}_j + m^{}_a$) are essentially invisible.

\subsection{Invisible Decays of GRB Neutrinos}
\label{subsec:invisible_decay_formalism}
With the effective interaction in Eq.~(\ref{eq:Lagint}), one can directly calculate the decay rate of $\nu_i^{}\to \nu_j^{}+a$. If massive neutrinos are Dirac particles, they can decay either to active left-handed neutrinos or to sterile right-handed neutrinos. On the other hand, if massive neutrinos are Majorana particles, both left-handed and right-handed neutrinos in the final states are active, where the latter ones actually correspond to the SM antineutrinos. Since we are interested in the ALPs produced in neutrino decays, the total decay rates in both channels of left-handed and right-handed neutrinos in the final state are relevant. For definiteness, massive neutrinos are assumed to be Majorana particles, and the total decay rate is given by~\cite{Kim:1990km,Funcke:2019grs}
\begin{eqnarray}\label{eq:decay_rate}
\Gamma\left(\nu_i^{} \to \nu_j^{}+a\right) = \frac{\left({\rm Re}\, y^{}_{ij}\right)^2 m_i^{}}{4 \pi}\left[\left(1 - \frac{m_j^{}}{m_i^{}}\right)^2 - \frac{m_a^2}{m_i^2}\right] \sqrt{\left(1 + \frac{m_j^{2}}{m_i^{2}} - \frac{m_a^2}{m_i^2}\right)^2 - \frac{4 m_j^2}{m_i^2}}\;,
\end{eqnarray}
where the decay rate of $\nu^{}_i \to \nu^{}_j + a$ and that of $\nu^{}_i \to \overline{\nu}^{}_j + a$ are summed up. In the case of Dirac neutrinos, the total rate in Eq.~(\ref{eq:decay_rate}) will be divided by a factor of four and the coupling ${\rm Re}\, y_{ij}^{}$ therein should be replaced by $\left|y_{ij}^{}\right|$~\cite{Funcke:2019grs}. Note that the decay rate of antineutrinos remains the same as in Eq.~(\ref{eq:decay_rate}), which should also be taken into account for the production of ALPs.

Suppose that neutrino mass ordering is normal, i.e., $m_1^{}<m_2^{}<m_3^{}$, and the lightest neutrino $\nu^{}_1$ is absolutely stable. Then the lifetimes of $\nu^{}_2$ and $\nu^{}_3$ are given by
\begin{eqnarray}
    \tau_2^{-1}&\equiv&\Gamma\left(\nu_2^{}\to \nu_1^{}+a\right)\;,\nonumber\\
    \label{eq:decay_rate3}
    \tau_3^{-1}&\equiv&\Gamma\left(\nu_3^{}\to \nu_1^{}+a\right)+\Gamma\left(\nu_3^{}\to \nu_2^{}+a\right)\;.
\end{eqnarray}
Given an extremely small mass $m^{}_a$ (e.g., $m^{}_a \approx 10^{-10}~{\rm eV}$), one can easily observe from Eq.~(\ref{eq:decay_rate}) that $\Gamma(\nu^{}_i \to \nu^{}_j + a) \propto (m^{}_i - m^{}_j)^3(m^{}_i + m^{}_j)/m^3_i$, implying that the decay rate crucially depends on the neutrino mass spectrum. In this work, we take the lightest neutrino to be massless, namely $m^{}_1 = 0$, and thus other two neutrino masses $m_2^{} \approx 8.61~{\rm meV} $ and  $m_3^{} \approx 50.1~{\rm meV} $ can be determined from current neutrino oscillation data~\cite{Esteban:2020cvm}. In this subsection, we consider the invisible decays of GRB neutrinos, for which only the combinations $\tau^{}_2/m^{}_2$ and $\tau^{}_3/m^{}_3$ will be relevant.

During the propagation of neutrinos over the cosmological distance, they will decay into ALPs and the number $N^{}_i(z)$ of neutrino mass eigenstate $\nu^{}_i$ changes with respect to the redshift $z$ as
\begin{eqnarray}
	\label{eq:diffnumber}
\frac{1}{N_i^{}(z)}\frac{{\rm d}N_i^{}(z)}{{\rm d}z}=-\frac{1}{\lambda_i^{}(z)}\frac{{\rm d}x(z)}{{\rm d}z}\;,
\end{eqnarray}
where $\lambda_i^{}(z)=(1+z)E_\nu^{}\tau_i^{}/m_i^{}$ is the decay length of $\nu_i^{}$ at redshift $z$ and $E_\nu^{}$ is the neutrino energy at $z=0$. Here $x(z)$ is the distance that $\nu_i^{}$ has traveled from the GRB (at $z_0^{}=0.15$) to redshift $z$ (i.e., the light-travel distance)~\cite{Baerwald:2012kc}. In a flat $\Lambda$CDM cosmology, $x(z)$ is given by
\begin{eqnarray}
	\label{eq:distance}
x(z)=\frac{c}{H_0^{}}\int_{z}^{z_0^{}}\frac{{\rm d}z'}{1+z'}\frac{1}{\sqrt{\Omega_m^{}\left(1+z'\right)_{}^3+\Omega_\Lambda^{}}}\;,
\end{eqnarray}
where $c/H_0^{}\approx 4446~{\rm Mpc}$ is the Hubble length, $\Omega_m^{}\approx 0.315$ and $\Omega_\Lambda^{}\approx 0.685$ are the energy-density fractions of matter and dark energy today~\cite{ParticleDataGroup:2022pth}. Note that $\ell_0^{}\equiv x(0)\approx 600~{\rm Mpc}$ corresponds to the physical distance between the GRB source and the detector. Substituting Eq.~(\ref{eq:distance}) into Eq.~(\ref{eq:diffnumber}), one obtains the ratio between the number of $\nu_i^{}$ at redshift $z$ and that at the GRB~\cite{Baerwald:2012kc}
\begin{eqnarray}
\frac{N_i^{}(z)}{N_i^{}(z_0^{})}={\rm exp}\left[-\frac{m_i^{}}{\tau_i^{}E_\nu^{}}\ell_{\rm eff}^{}(z)\right]\;,
\end{eqnarray}
where the effective distance is defined as
\begin{eqnarray}
\ell_{\rm eff}^{}(z)\equiv\frac{c}{H_0^{}}\int_{z}^{z_0^{}}\frac{{\rm d}z'}{\left(1+z'\right)_{}^2}\frac{1}{\sqrt{\Omega_m^{}\left(1+z'\right)_{}^3+\Omega_\Lambda^{}}}\;.
\end{eqnarray}
Therefore, denoting the $\nu^{}_i$ flux at the source as $\phi^{}_i$ and noticing that the survival probability of GRB neutrinos is $N^{}_i(0)/N^{}_i(z^{}_0)$, one can find the ALP flux from $\nu_i^{}$ decays\footnote{Notice that we have neglected the contribution to the ALP flux from the secondary neutrino $\nu^{}_2$, which comes from $\nu_3^{}$ decays (i.e., $\nu_3^{}\to \nu_2^{}+a$ and $\nu^{}_2 \to \nu_1^{}+a$). This is reasonable because the lifetime of $\nu_3^{}$ is very long anyway, as we shall explain later.}
\begin{eqnarray}
	\phi_a^i=\phi_i^{}\left[1-\frac{N_i^{}(0)}{N_i^{}(z_0^{})}\right]=\phi_i^{}\left[1-{\rm exp}\left(-\frac{m_i^{}\ell_{\rm eff}^{}}{\tau_i^{}E_\nu^{}}\right)\right]\;,
\end{eqnarray}
with the effective distance $\ell_{\rm eff}^{}\equiv \ell_{\rm eff}^{}(0)\approx 560~{\rm Mpc}$. In consideration of the initial fluxes of GRB neutrinos, the total flux of ALPs turns out to be
\begin{eqnarray}
	\label{eq:transitionrate}
\phi_a^{}=\sum_{i=2,3}^{}\phi_a^{i}=\sum_{i=2,3}^{}\left[1-{\rm exp}\left(-\frac{m_i^{}\ell_{\rm eff}^{}}{\tau_i^{}E_\nu^{}}\right)\right] \sum_{\alpha}\left|U_{\alpha i}^{}\right|_{}^2\phi_{\nu_\alpha^{}}^{}\;,
\end{eqnarray}
where $\phi_{\nu_\alpha^{}}^{}$ stands for the $\nu_\alpha^{}$ flux. In the left and middle panels of Fig.~\ref{fig:Enu}, the fluxes of $\phi_{\nu_e^{}}^{}$ and $\phi_{\nu_\mu^{}}^{}$ at the source have been shown for typical input parameters of the GRB model. The flux of ALPs can also be written as $\phi_a^{} = P_{\nu_e^{}a}^{}\,\phi_{\nu_e^{}}^{} + P_{\nu_\mu^{}a}^{}\,\phi_{\nu_\mu^{}}^{}$, where the probability for the ALP production from $\nu^{}_\alpha$ (for $\alpha = e, \mu$) is given by
\begin{eqnarray}
		\label{eq:Pnua}
P_{\nu_\alpha^{}a}^{}\left(E_\nu^{}\right)=\sum_{i=2,3}^{}\left[1-{\rm exp}\left(-\frac{m_i^{}\ell_{\rm eff}^{}}{\tau_i^{}E_\nu^{}}\right)\right]\left|U_{\alpha i}^{}\right|_{}^2\;.
\end{eqnarray}
To be more explicit, using the standard parametrization of the leptonic flavor mixing matrix~\cite{ParticleDataGroup:2022pth}, we can get
\begin{eqnarray}
	\label{eq:Pnua_analy}
P_{\nu_e^{}a}^{}\left(E_\nu^{}\right)&=&\left[1-{\rm exp}\left(-\frac{m_2^{}\ell_{\rm eff}^{}}{\tau_2^{}E_\nu^{}}\right)\right]c_{13}^2s_{12}^2+\left[1-{\rm exp}\left(-\frac{m_3^{}\ell_{\rm eff}^{}}{\tau_3^{}E_\nu^{}}\right)\right]s_{13}^2\;,\nonumber\\
P_{\nu_\mu^{}a}^{}\left(E_\nu^{}\right)&=&\left[1-{\rm exp}\left(-\frac{m_2^{}\ell_{\rm eff}^{}}{\tau_2^{}E_\nu^{}}\right)\right]\left(
c_{12}^2 c_{23}^2+s_{12}^2 s_{13}^2 s_{23}^2-2c_{12}^{}c_{23}^{}s_{12}^{}s_{13}^{}s_{23}^{}\cos\delta
\right)\nonumber\\
&+&\left[1-{\rm exp}\left(-\frac{m_3^{}\ell_{\rm eff}^{}}{\tau_3^{}E_\nu^{}}\right)\right]c_{13}^2s_{23}^2\;,
\end{eqnarray}
where $c_{ij}^{}\equiv\cos\theta_{ij}^{}$ and $s_{ij}^{}\equiv \sin\theta_{ij}^{}$ have been defined with $\theta_{ij}^{}$ (for $ij = 12, 13, 23$) being the leptonic flavor mixing angles and $\delta$ being the Dirac CP-violating phase.

In the literature, the visible or invisible decays of high-energy neutrinos have been extensively studied~\cite{Pakvasa:1999ta,Beacom:2002vi,Beacom:2003zg,Meloni:2006gv,Maltoni:2008jr,Xing:2008fg,Choubey:2009jq,Dorame:2013lka,Pagliaroli:2015rca,Bustamante:2016ciw,Denton:2018aml,Palladino:2019pid,Bustamante:2020niz,Abdullahi:2020rge,Shoemaker:2015qul}. Different from previous investigations, in which high-energy neutrinos after decays and their detection at neutrino telescopes are of main interest, we focus on the invisible ALPs in the final state of neutrino decays. In our case, the flux of high-energy muon neutrinos from GRB221009A has been required to satisfy the IceCube bound, as indicated in the middle panel of Fig.~\ref{fig:Enu}, so this is also true after neutrino decays.

\subsection{Constraints on Neutrino Lifetimes}
\label{subsec:invisible_decay_numerical}
From Eq.~(\ref{eq:Pnua_analy}) it can be seen that the probability for the production of ALPs from neutrino decays depends on the neutrino energy $E_\nu^{}$ and neutrino lifetimes, i.e., $\tau_2^{}/m_2^{}$ and $\tau_3^{}/m_3^{}$. In particular, all these parameters appear in the exponential, so the probability decreases rapidly as the neutrino energy and lifetime increase. For this reason, it is important to examine the present constraints on neutrino lifetimes.

Undoubtedly, invisible neutrino decays receive quite a number of constraints from terrestrial neutrino experiments, astrophysics and cosmology. Among these constraints, the most stringent bound on the neutrino lifetime comes from cosmology~\cite{Barenboim:2020vrr}
\begin{eqnarray}
	\label{eq:constraint}
	\tau_\nu^{} > 4 \times 10^5~{\rm s}\left(\frac{m_\nu^{}}{50~{\rm meV}}\right)^5\;,
\end{eqnarray}
where $m_\nu^{}$ denotes generally the mass of a heavier active neutrino that can decay into a lighter neutrino and a massless (pseudo-)scalar particle~\cite{Barenboim:2020vrr}.\footnote{The constraint in Eq.~(\ref{eq:constraint}) is actually valid for the massless daughter neutrino. If the mass of the daughter neutrino is taken into account, a new phase-space factor comes into play and weakens the constraint on the neutrino lifetime~\cite{Chen:2022idm}. The authors are grateful to Prof. Yvonne Y. Y. Wong for helpful communications about this point.} The bound in Eq.~(\ref{eq:constraint}) is applicable to the case where relativistic neutrinos decay substantially. For the invisible decays of non-relativistic neutrinos, the bound needs to be revised in a systematic way~\cite{Barenboim:2020vrr}. Whether neutrinos decay relativistically or non-relativistically depends on the neutrino mass and the temperature of the Universe when neutrino decay rate exceeds the Hubble expansion rate. Other constraints on invisible neutrino decay from Supernova 1987A, Big Bang Nucleosynthesis, solar neutrinos, atmospheric neutrinos, and high-energy astrophysical neutrinos measured at IceCube are less restrictive than the cosmological constraint (see, e.g., Refs.~\cite{Escudero:2019gfk,FrancoAbellan:2021hdb} and references therein). 

As has been mentioned before, the normal neutrino mass ordering with $m_1^{}=0$, $m_2^{} \approx 8.61~{\rm meV} $ and  $m_3^{} \approx 50.1~{\rm meV}$ will be considered. With the help of Eqs.~(\ref{eq:decay_rate}) and (\ref{eq:decay_rate3}), as well as the condition $m^{}_2, m^{}_3 \gg m^{}_a$, the lifetimes of two unstable neutrinos can be estimated as
\begin{eqnarray}
	\frac{\tau_2^{}}{m_2^{}}  & \approx & 8\times 10^6 \left(\frac{{\rm Re}\,y^{}_{21}}{1.25 \times 10^{-7}}\right)^{-2}\left(\frac{m_2^{}}{50~{\rm meV}}\right)^4 ~{\rm s}~{\rm eV}^{-1} \;, \label{eq:lifetime_2} \\
	\frac{\tau_3^{}}{m_3^{}}  & \approx & 8\times 10^6 \left(\frac{{\rm Re}\,y^{}_{31}}{6.39 \times 10^{-10}}\right)^{-2}\left(\frac{m_3^{}}{50~{\rm meV}}\right)^4~{\rm s}~{\rm eV}^{-1} \;. \label{eq:lifetime_3}
\end{eqnarray}
Therefore, the cosmological constraint in Eq.~(\ref{eq:constraint}) can be fulfilled when the couplings ${\rm Re}\,y_{21}^{}\sim 10^{-7}_{}$ and ${\rm Re}\,y_{31}^{} \sim {\rm Re}\,y_{32}^{} \sim 10^{-10}_{}$ are chosen. 

\begin{figure}[t]
	\centering
	\includegraphics[scale=0.75]{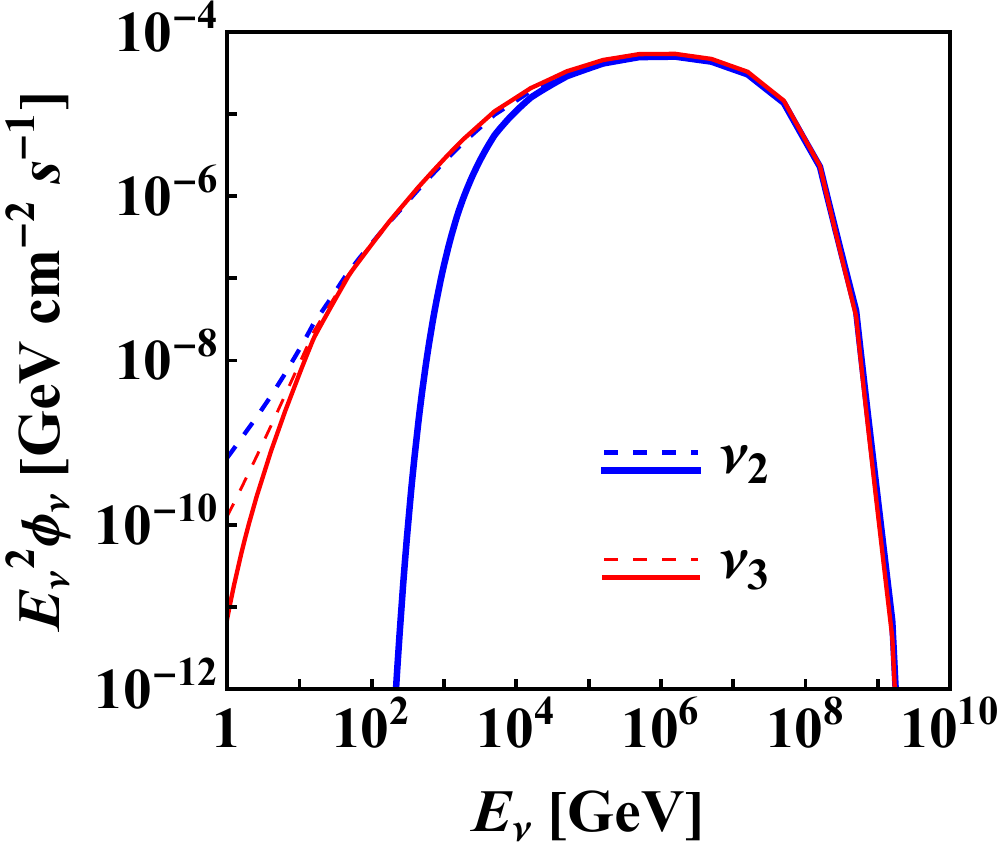}\qquad
    \includegraphics[scale=0.75]{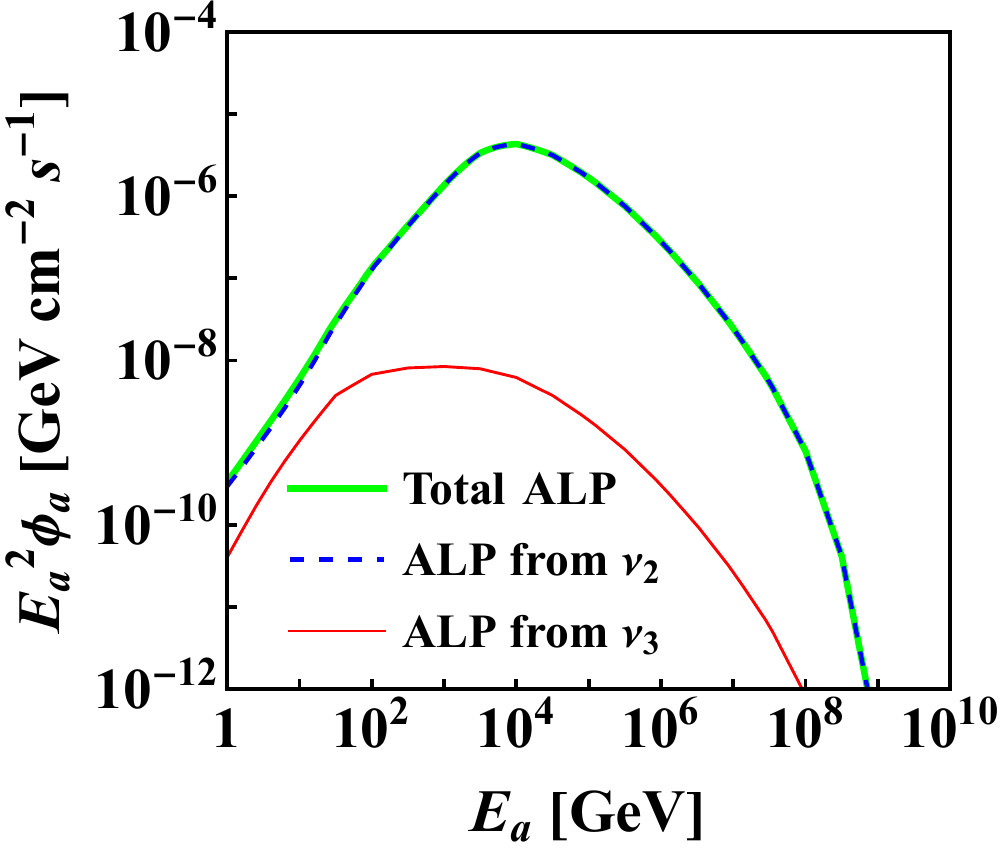}
	\caption{{\it Left}: The fluxes of neutrino mass eigenstates $\nu_2^{}$ and $\nu_3^{}$ at the source (dashed curves) and in the Milky Way (solid curves), where the neutrino lifetimes $\tau_2^{}/m_2^{} = 10^4_{}~{\rm s}~{\rm eV}^{-1}$ and $\tau_3^{}/m_3^{}= 10^7~{\rm s}~{\rm eV}^{-1}$ are taken. {\it Right}: The fluxes of ALPs in the Milky Way from neutrino decays. The total flux is shown as green solid curve, which is obtained by adding up the contribution from the decay of $\nu_2^{}$ (blue dashed curve) and that of $\nu_3^{}$ (red solid curve).} 
	\label{fig:nuALP_spectrum}
\end{figure}

For a ballpark feeling about the ALP production from neutrino decays, we take $E_\nu^{} = 1~{\rm TeV}$, $\tau_2^{}/m_2^{}= 10^4_{}~{\rm s}~{\rm eV}^{-1}$ and $\tau_3^{}/m_3^{} = 10^7_{}~{\rm s}~{\rm eV}^{-1}$, which are well compatible with the cosmological constraint in Eq.~(\ref{eq:constraint}). In addition, we use the best-fit values of neutrino oscillation parameters in the case of normal mass ordering from Ref.~\cite{Esteban:2020cvm}: $s_{12}^2=0.304$, $s_{23}^2=0.573$, $s_{13}^2=0.02219$ and $\delta=197^{\circ}_{}$. Then, one can evaluate the production probability of the ALPs in Eq.~(\ref{eq:Pnua_analy}) as
\begin{eqnarray}
	\label{eq:Pnua_benchmark}
	P_{\nu_e^{}a }^{}\left(E_\nu^{}=1~{\rm TeV}\right)\approx 0.30\;,\qquad P_{\nu_\mu^{}a }^{}\left(E_\nu^{}=1~{\rm TeV}\right) \approx 0.37 \;,
\end{eqnarray}
which are sizable enough. For further illustration, we show the fluxes of $\nu_2^{}$ and $\nu_3^{}$ both at the source and in the MW in Fig.~\ref{fig:nuALP_spectrum}, as well as the fluxes of ALPs in the MW from neutrino decays. Some comments are helpful. First, as the lightest neutrino $\nu^{}_1$ is absolutely stable and it does not contribute to the ALP production, its flux is not shown in Fig.~\ref{fig:nuALP_spectrum}. On the other hand, the fluxes of $\nu^{}_2$ and $\nu^{}_3$ at the source (blue and red dashed curves) have been obtained by projecting the fluxes of neutrino flavor states in Fig.~\ref{fig:Enu} into the mass basis. Second, the fluxes of $\nu^{}_2$ and $\nu^{}_3$ in the MW (blue and red solid curves) have been modified by invisible decays. However, it should be mentioned that the production of $\nu^{}_2$ from $\nu^{}_3$ decays has not been taken into account for simplicity. A more complete treatment is to solve the coupled differential equations for neutrino fluxes by including both decays and production. Third, as one can observe from the right panel of Fig.~\ref{fig:nuALP_spectrum}, the main contribution to the ALP production is from $\nu^{}_2$ decays, since the lifetime of $\nu^{}_2$ is much shorter than that of $\nu^{}_3$. It is worthwhile to note that neutrinos with energies higher than $100~{\rm TeV}$ have rarely decayed when arriving in the MW.

\section{ALP-Photon Conversion}
\label{sec:ALP-photon_converison}

Because of the interaction with two photons, as given in Eq.~(\ref{eq:Laga}), the ALPs may convert to photons in the presence of external magnetic fields~\cite{Raffelt:1987im} (see Refs.~\cite{Jaeckel:2010ni, Marsh:2015xka, Irastorza:2018dyq, DiLuzio:2020wdo} for some recent reviews of ALPs and their detection). As for photons, the dispersion relation in media is of crucial importance. First, radiative corrections in quantum electrodynamics (QED) to the dispersion can be described by the effective Euler--Heisenberg Lagrangian\footnote{In the calculation of the ALP-photon conversion probability, we use the rationalized natural units where $c = \hbar = 1$ and the fine-structure constant $\alpha$ is related to the charge of the electron $e$ by $\alpha=e_{}^{2}/(4\pi)$. In such a system of units~\cite{raffelt1996stars}, we have $1~\mu{\rm G} = 1.95\times 10_{}^{-26}~{\rm GeV}_{}^2$.}
\begin{eqnarray} \label{eq:ALP-photon_mixing_Lagrangian}
    {\cal L}^{}_\gamma = - \frac{1}{4} F_{\mu\nu}^{} F^{\mu\nu}_{} + \frac{\alpha^2_{}}{90 m_e^4} \left[\left(F_{\mu\nu}^{} F^{\mu\nu}_{}\right)^2 + \frac{7}{4} \left(F_{\mu\nu}^{} \tilde{F}^{\mu\nu}_{}\right)^2\right] \;,
\end{eqnarray}
where $\alpha \approx 1/137$ is the electromagnetic fine-structure constant and $m_e^{} \approx 0.511~{\rm MeV}$ is the electron mass. The effective interactions between photons mediated by a pair of virtual electrons and positrons, as indicated by two terms in the square parentheses in Eq.~(\ref{eq:ALP-photon_mixing_Lagrangian}), contribute to two polarization states of the photon differently. If the ALP travels along the $z$-axis in the transverse magnetic field $B^{}_e$, the amplitudes of two polarization states $\{A^{}_\perp, A^{}_\parallel\}$ of the photon and that $a$ of the ALP evolve according to~\cite{Raffelt:1987im}
\begin{eqnarray} \label{eq:wave_equation}
    \left[ \omega +  \begin{pmatrix} \Delta_\perp^{} & 0 & 0 \\ 0 & \Delta_\parallel^{} & \Delta_g^{} \\ 0 & \Delta_g^{} & \Delta_a^{} \end{pmatrix} + {\rm i} \partial_z \right]\begin{pmatrix} A_\perp^{} \\ A^{}_\parallel \\ a \end{pmatrix}=0\;,
\end{eqnarray}
where $\omega$ is the wave frequency of the ALP and the converted photon, and $A^{}_\perp$ (or $A^{}_\parallel$) denotes the amplitude perpendicular (or parallel) to the transverse magnetic field. The corrections to the photon dispersion relation are denoted by
\begin{eqnarray}
	\Delta_\parallel^{} = \Delta_{\rm pl}^{} + \frac{7}{2} \Delta_{\rm QED}^{} + \Delta_{\rm CMB}\;, \qquad
	\Delta_\perp^{} = \Delta_{\rm pl}^{} + 2 \Delta_{\rm QED}^{} + \Delta_{\rm CMB}\;,
 \label{eq:photon_hamiltonian_long}
\end{eqnarray}
where $\Delta^{}_{\rm QED} = 4\alpha^2_{} B_e^2 \omega / (45  m_e^4)$ arises from the Euler--Heisenberg interaction term~\cite{Adler:1971wn,Raffelt:1987im}. Then, the refractive index of photons in the plasma is given by $\chi^{}_{\rm pl} = -\omega^2_{\rm pl}/(2\omega^2)$, where $\omega^{2}_{\rm pl} \equiv 4\pi \alpha n^{}_e/m^{}_e$ is the plasma frequency with $n^{}_e$ being the electron number density. In the presence of CMB, the refractive index turns out to be $\chi^{}_{\rm CMB} = 44\alpha^2 \rho^{}_{\rm CMB}/(135 m^4_e)$, where the electromagnetic energy density of CMB is $\rho^{}_{\rm CMB} = \pi^2 T^4/15$ with today's temperature $T \approx 2.726~{\rm K}$. These refractive indices of photons are related to the corrections in Eq.~(\ref{eq:photon_hamiltonian_long}) via $\Delta^{}_{\rm pl} \equiv \chi^{}_{\rm pl} \omega$ and $\Delta^{}_{\rm CMB} \equiv \chi^{}_{\rm CMB} \omega$.

As for the ALP, the diagonal element in the evolution matrix in Eq.~(\ref{eq:ALP-photon_mixing_Lagrangian}) is $\Delta_a^{} = -m_a^2 / (2 \omega)$. The off-diagonal element $\Delta_g^{} = g_{a\gamma}^{} B_e^{} /2$ accounts for the interaction of the ALP with the transverse polarization of photons and the external transverse magnetic field $B_e^{} $. Since the evolution of the amplitude $A^{}_\perp$ completely decouples, as indicated in Eq.~(\ref{eq:wave_equation}), we are left with a two-dimensional evolution matrix. After diagonalizing the matrix, one can find the mixing angle and two eigenvalues
\begin{eqnarray}\label{eq:mixing_angle_eigenvalues}
    \tan2\theta = \frac{2 \Delta_g^{}}{\Delta_\parallel - \Delta_a^{}}\;, \quad \lambda_\pm^{} = \frac{\Delta^{}_\parallel + \Delta^{}_a}{2} \pm \frac{\sqrt{\left(\Delta^{}_\parallel-\Delta^{}_a\right)^2 + 4 \Delta_g^2}}{2}\;, 
\end{eqnarray}
and calculate the probability of ALP-photon conversion~\cite{Dobrynina:2014qba}
\begin{eqnarray}\label{eq:probability}
    P_{a\gamma}^{} = \sin^2 \left(2\theta \right) \sin^2\left(\frac{\pi L}{L_{\rm osc}^{}}\right)\; .
\end{eqnarray}
In Eq.~(\ref{eq:probability}), $L$ is the traveling distance of the ALP in the magnetic field and the oscillation length is defined as $L_{\rm osc} = 2\pi/(\lambda^{}_{+} - \lambda^{}_{-}) = 2\pi/[(\Delta^{}_\parallel-\Delta^{}_a)^2 + 4 \Delta_g^2]^{1/2}$. It is straightforward to observe from Eq.~(\ref{eq:mixing_angle_eigenvalues}) that the mixing angle $\theta$ depends sensitively on the ALP-photon coupling constant $g^{}_{a\gamma}$ and the magnetic field $B^{}_e$ in $\Delta^{}_g$, but the photon dispersion and the ALP mass in the denominator could also be important. The interplay between these parameters renders the phenomena of the ALP-photon conversion to be very interesting.

\begin{figure}[t]
	\centering
	\includegraphics[scale=0.6]{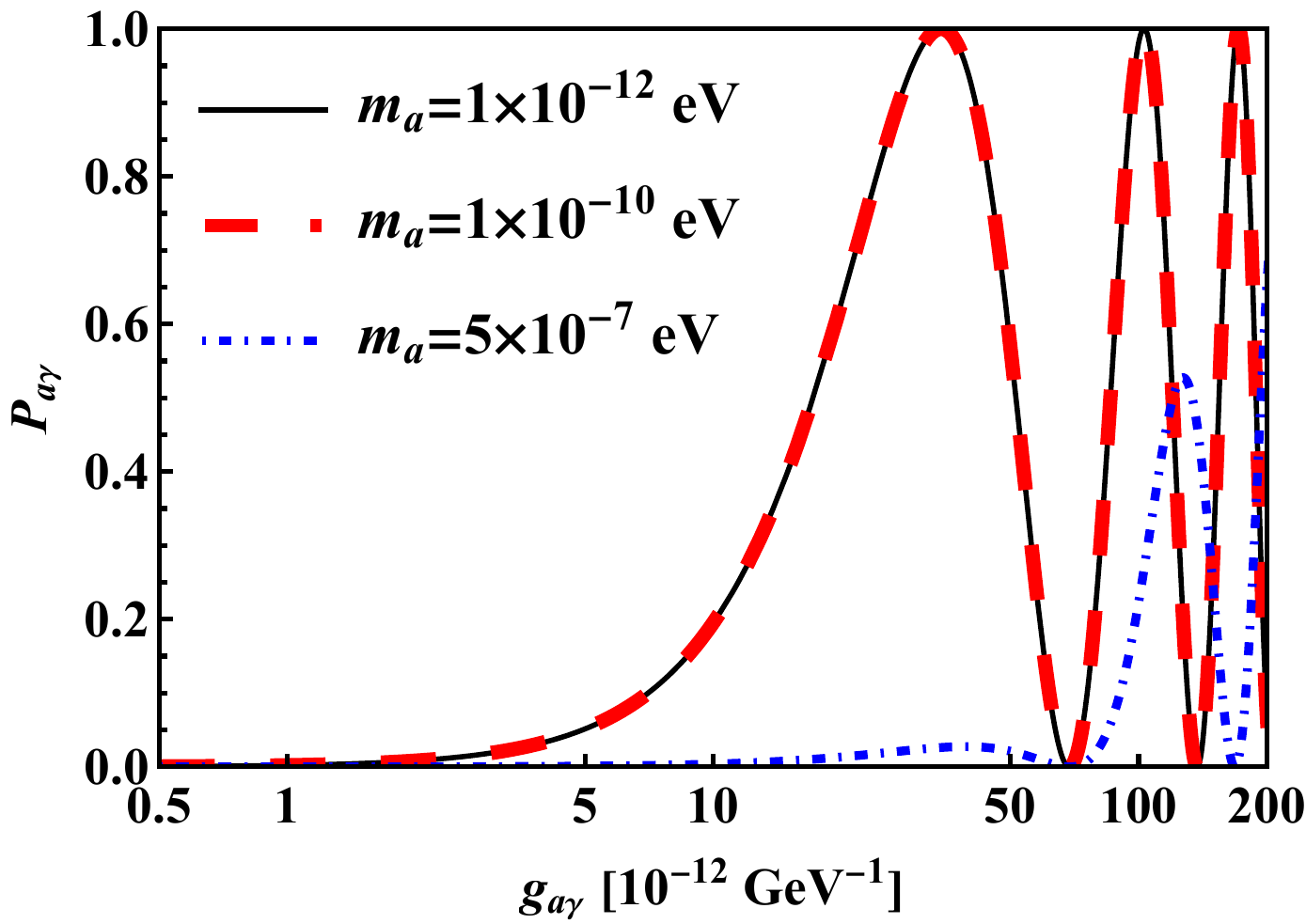}\ 
        \includegraphics[scale=0.63]{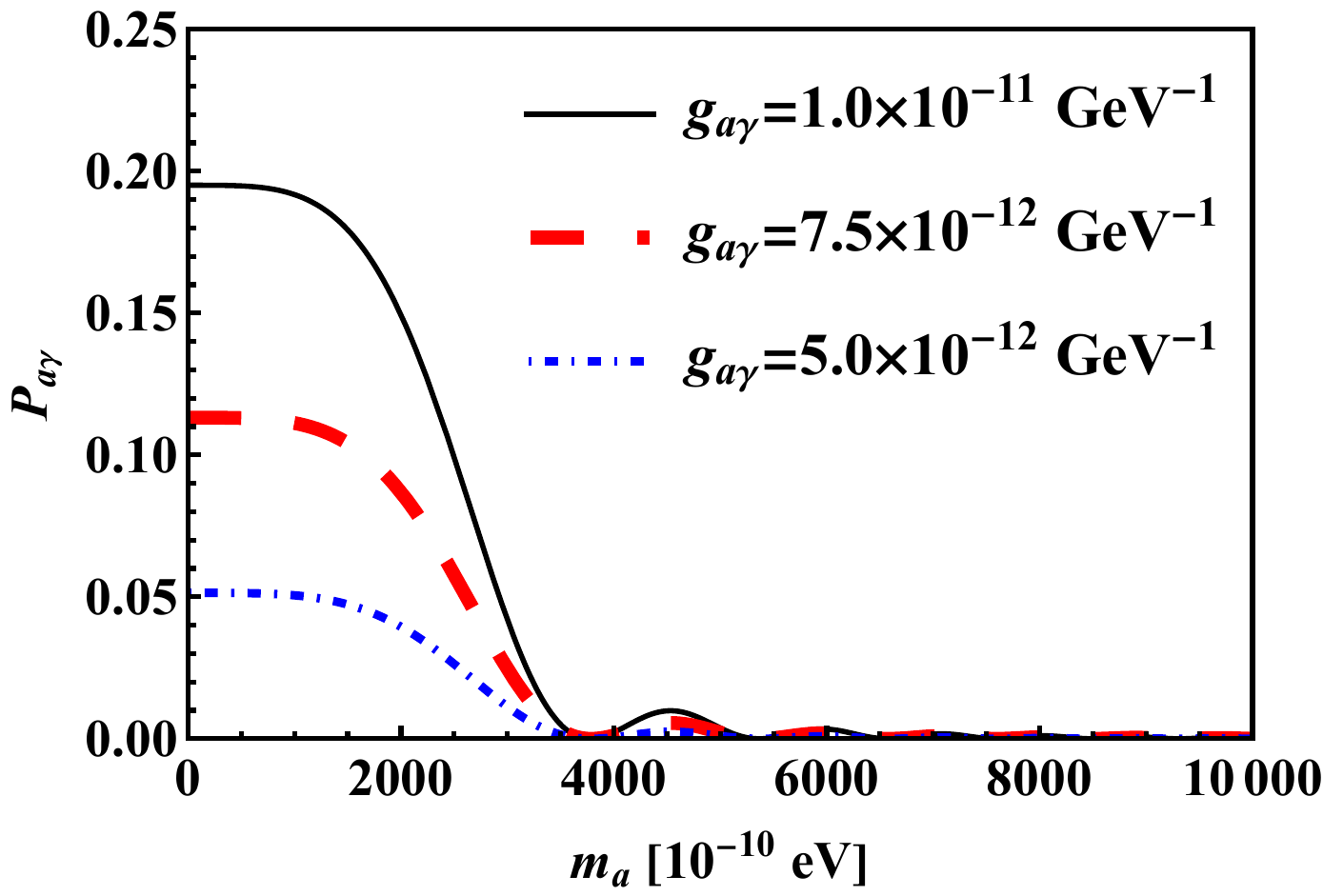}
    \vspace{-0.2cm}
	\caption{The ALP-photon conversion probability versus the ALP-photon coupling $g_{a\gamma}^{}$ (left) and the ALP mass $m_a^{}$ (right), where the ALP energy is fixed at $\omega=18~{\rm TeV}$. In the left panel, the ALP mass is taken to be $m_a^{}=10_{}^{-12}~{\rm eV}$ (black solid curve), $10_{}^{-10}~{\rm eV}$ (red dashed curve) and $5\times 10_{}^{-7}~{\rm eV}$ (blue dot-dashed curve). In the right panel, the ALP-photon coupling is taken to be $g_{a\gamma} = 10^{-11}~{\rm GeV}^{-1}$ (black solid curve), $7.5 \times 10^{-12}~{\rm GeV}^{-1}$ (red dashed curve) and $5 \times 10^{-12}~{\rm GeV}^{-1}$ (blue dot-dashed curve).}
	\label{fig:prob_g_ma}
\end{figure}
In Fig.~\ref{fig:prob_g_ma}, we show the conversion probability $P_{a\gamma}^{}$ in Eq.~(\ref{eq:probability}) for the ALP energy at $\omega=18~{\rm TeV}$ by varying either the ALP-photon coupling $g_{a\gamma}^{}$ or the ALP mass $m_a^{}$. 
In our numerical calculations, we focus on the ALP-photon conversion within the MW, and thus set the average strength of the galactic magnetic field to be $B_e^{}\approx 3~\mu{\rm G}$ and the electron density of the MW to be $n_e^{} \approx 10^{-3}~{\rm cm}^{-3} $~\cite{Dobrynina:2014qba}. In addition, a typical value of $L\approx 10~{\rm kpc}$ for the traveling distance of ALPs in the MW is assumed.
Two helpful comments are in order:
\begin{itemize}
\item First, in the limit of small couplings, i.e., $2\Delta_g^{}\ll (\Delta^{}_\parallel-\Delta_a^{})$, the conversion probability approximates to  
\begin{eqnarray}
	P_{a\gamma}^{} \approx \frac{g_{a\gamma}^2 B_e^2}{\left(\Delta^{}_\parallel-\Delta_a^{} \right)_{}^2}\sin_{}^2\left[\left( \Delta^{}_\parallel-\Delta_a^{} \right)\frac{L}{2}\right]\; ,
\end{eqnarray}
where its overall magnitude is proportional to $g_{a\gamma}^2$ and will be highly suppressed. Such suppression is clear from the left panel of Fig.~\ref{fig:prob_g_ma}. Meanwhile, the oscillation length $L^{}_{\rm osc} \approx 2\pi/|\Delta^{}_\parallel-\Delta_a^{}|$ is independent of the coupling. On the other hand, if the coupling is large, i.e., $2\Delta_g^{}\gg(\Delta^{}_\parallel-\Delta_a^{})$, the mixing angle becomes maximal (namely, $\theta\to \pi/4$) and the probability oscillates with an increasing coupling
\begin{eqnarray}
P_{a\gamma}^{}=\sin^2_{}\left(\frac{g_{a\gamma}^{} B_e^{}}{2}L\right)\;,
\end{eqnarray}
as shown in the right part of the left panel of Fig.~\ref{fig:prob_g_ma}. The critical value of the coupling, with which the magnitudes of $2\Delta^{}_g$ and $\Delta^{}_\parallel-\Delta_a^{}$ are comparable, can be estimated as $g_{\rm c}^{} = ( \Delta^{}_\parallel-\Delta_a^{})/B_e^{}\approx 0.41\times 10^{-12}_{}~{\rm GeV^{-1}_{}}$.
    
\item Second, for a small ALP mass, we have $\left|\Delta_a^{}\right|\ll \Delta_{\parallel}^{}$, and thus the conversion probability is almost independent of $m_a^{}$. For a large ALP mass, i.e., $\left|\Delta_a^{}\right|\gg \Delta_{\parallel}^{}$, the probability would be proportional to $1/m_a^4$ and decrease quickly as $m_a^{}$ increases. In this work, we are interested in the ultra-light ALP with $m_a^{}\sim 10^{-10}_{}~{\rm eV}$,\footnote{Such a light ALP has a lifetime much longer than the age of the Universe, thus has no influence on the cosmic evolution~\cite{Cadamuro:2010cz,Cadamuro:2011fd}.} so the conversion probability is insensitive to the ALP mass, as is shown in the right panel of Fig.~\ref{fig:prob_g_ma}. This can also be seen from the left panel of Fig.~\ref{fig:prob_g_ma}, where the black solid curve for $m_a^{} = 10^{-12}~{\rm eV}$ coincides with the red dashed one for $m_a^{} = 10^{-10}~{\rm eV}$, while for heavier ALP of mass $m_a^{} = 5\times 10^{-7}~{\rm eV} $ (blue dot-dashed curve) the probability is suppressed. 
\end{itemize}
As a benchmark value, for $m_a^{}=10_{}^{-10}~{\rm eV}$ and $g_{a \gamma}^{}=5\times 10_{}^{-12}~ {\rm GeV}_{}^{-1}$, one obtains the ALP-photon conversion probability $P_{a\gamma}^{} \left(\omega = 18~{\rm TeV}\right) \approx 5.15 \times 10^{-2}$ in the MW.

Finally, during the propagation of the ALPs in the extragalactic magnetic field $B_{\rm ext}^{}$, they may also be converted to photons. The exact information about the extragalactic magnetic field is still lacking, and the present bound $10^{-16}_{}~{\rm G} \lesssim B_{\rm ext}^{}\lesssim {\cal O}({\rm nG})$ allows it to vary by many orders of magnitude~~\cite{Neronov:2010gir,Pshirkov:2015tua}.
To examine whether the extragalactic magnetic field can play any role in our scenario, we evaluate the relevant quantities in the evolution matrix as follows~\cite{Dobrynina:2014qba}
\begin{eqnarray}\label{eq:Delta_g/a}
\Delta_a &=& - \frac{m_a^2}{2\omega} \approx -4.34 \times 10^{-8}\left(\frac{m_a^{}}{10^{-10}~{\rm eV}}\right)^2 \left(\frac{\omega}{18~{\rm TeV}}\right)^{-1}~{\rm kpc}^{-1}\;, \nonumber \\
\Delta_g &=& +\frac{g_{a\gamma}^{} B_e^{}}{2} \approx +2.29 \times 10^{-2}\left(\frac{g^{}_{a\gamma}}{5 \times 10^{-12}~{\rm GeV}^{-1}}\right)\left(\frac{B^{}_{e}}{3~\mu{\rm G}}\right)~{\rm kpc}^{-1}\;,  \nonumber \\
\Delta_{\rm pl}^{} &=& -\frac{2 \pi \alpha n_e}{m_e \omega} \approx -5.98 \times 10^{-12} \left(\frac{n_e^{}}{10^{-3}~{\rm cm}^{-3}}\right) \left(\frac{\omega}{18~{\rm TeV}}\right)^{-1}~{\rm kpc}^{-1}\;, \nonumber \\
\Delta_{\rm QED}^{} &=& +\frac{4\alpha^2 B_e^2 \omega}{45 m_e^4} \approx +6.69 \times 10^{-4} \left(\frac{B^{}_{e}}{3~\mu{\rm G}}\right)^2 \left(\frac{\omega}{18~{\rm TeV}}\right)~{\rm kpc}^{-1}\;, \nonumber \\
\Delta_{\rm CMB}^{} &=& +\chi_{\rm CMB}^{} \, \omega \approx +1.44 \times 10^{-3}\left(\frac{\omega}{18~{\rm TeV}}\right)~{\rm kpc}^{-1}\;, 
\end{eqnarray}
where the relevant parameters have been normalized to those in the MW~\cite{Pshirkov:2011um,Jansson:2012pc,Jansson:2012rt}. Even for the upper bound on the extragalactic magnetic field $B^{}_{\rm ext} \sim {\cal O}({\rm nG})$, one obtains $\Delta^{}_g \approx 10^{-5}~{\rm kpc}^{-1}$ and $\Delta^{}_{\parallel} - \Delta^{}_a \approx \Delta^{}_{\rm CMB} \approx 10^{-3}~{\rm kpc}^{-1}$, where tiny contributions from $\Delta^{}_{\rm pl}$, $\Delta^{}_{\rm QED}$ and $\Delta^{}_a$ are ignored. As a consequence of $\Delta^{}_g \ll \Delta^{}_{\parallel} - \Delta^{}_a$, the mixing angle $\theta$ turns to be small and the conversion probability will be highly suppressed. 
It has been demonstrated in Ref.~\cite{Galanti:2022pbg} that no substantial differences in the conversion probability would arise for high-energy photons with or without the inclusion of nano-Gauss $B_{\rm ext}^{}$. Therefore, we shall neglect the contribution from the extragalactic magnetic field in the subsequent discussions.

The ALP-photon conversion probability shown in Fig.~\ref{fig:prob_g_ma} is based on the assumption of a constant magnetic field, which should be compared with the analytical expression in Eq.~(\ref{eq:probability}). However, the magnetic field in the MW is not constant in reality. In this case, Eq.~(\ref{eq:probability}) is valid only in a small enough domain, where the magnetic field and the electron density are approximately constant. Then the final conversion probability from the source to the observer is proportional to the multiplication of the transfer matrices in each domain, while the latter can be obtained by numerically solving Eq.~(\ref{eq:wave_equation}) for constant magnetic fields. Therefore, the value of $P_{a\gamma}^{}$ depends on the magnetic field distribution in the MW. For the calculation of the event number of high-energy photons in the next section, we shall actually use {\tt gammaALPs}~\cite{Meyer:2021pbp}, an open-source {\tt Python} package, to compute the ALP-photon conversion probability in the MW. The package {\tt gammaALPs} adopts the widely-used models of the galactic magnetic field~\cite{Pshirkov:2011um,Jansson:2012pc,Jansson:2012rt} and allows for a more realistic calculation of $P_{a\gamma}^{}$.

\section{Gamma Rays at LHAASO}
\label{sec:photon_numbers}
Now that the flux of gamma rays converted from the ALPs in the MW and that of the gamma rays directly coming from the GRB are obtained, we can compute the expected number of VHE photons at LHAASO. First, we consider the photons from invisible neutrino decays. Within the energy range of $E_{\rm min}^{}<E_\gamma^{}<E_{\rm max}^{}$, where the upper and lower limits will be specified for a given detector, the event number reads
\begin{eqnarray}
	\label{eq:photon_number}
N_\gamma^{\nu}=
\Delta t \int_{E_{\rm min}^{}}^{E_{\rm max}^{}} \left[\phi_{\nu_e^{}}^{}\left(E_\gamma^{}\right)P_{\nu_e^{}a}^{}\left(E_\gamma^{}\right)+\phi_{\nu_\mu^{}}^{}\left(E_\gamma^{}\right)P_{\nu_\mu^{}a}^{}\left(E_\gamma^{}\right)\right] P_{a\gamma}^{}\left(E_\gamma^{}\right)A_{\rm eff}^{}\left(E_\gamma^{}\right){\rm d}E_{\gamma}^{}\;,
\end{eqnarray}
where $\Delta t=2000~{\rm s}$ is the observation time of LHAASO~\cite{GCN32677} and $A_{\rm eff}^{}$ is the effective area of the LHAASO detectors~\cite{Ma:2022aau}.\footnote{The effective area and energy resolution of the detector LHAASO-KM2A are taken from Fig.~2 of \cite{Ma:2022aau}, while those of another detector LHAASO-WCDA are taken from Fig.~26 of \cite{Ma:2022aau}.} The neutrino fluxes $\phi_{\nu_\alpha^{}}^{}$ (for $\alpha=e,\mu$) have been given in the left panel of Fig.~\ref{fig:Enu}, while the ALP production probabilities $P_{\nu_{\alpha}^{}a}^{}$ are calculated in Eq.~(\ref{eq:Pnua_analy}). Moreover, the {\tt Python} package {\tt gammaALPs}~\cite{Meyer:2021pbp} is used to compute the ALP-photon conversion probability $P_{a\gamma}^{}$ in the MW, where the galactic magnetic field model introduced in~\cite{Pshirkov:2011um} is taken as the input. We find that the actual value of $P_{a\gamma}^{}$ is smaller than that computed using an average constant magnetic field in Sec.~\ref{sec:ALP-photon_converison}. For example, when the ALP energy is taken to be 18~TeV, the conversion probability calculated from {\tt gammaALPs} is about $P(\omega=18~{\rm TeV})\approx 1.06\times 10_{}^{-2}$ for $m_a^{}=10_{}^{-10}$~{\rm eV} and $g_{a\gamma}^{}=5\times 10_{}^{-12}~{\rm GeV}_{}^{-1}$, which is about 1/5 of the result in a constant magnetic field $B_e^{}\approx 3~{\mu {\rm G}}$ calculated in Sec~\ref{sec:ALP-photon_converison}. Notice that the arguments in the neutrino fluxes and the conversion probabilities have been changed from $E_\nu^{}$ to $E_\gamma^{}$ with the substitution $E_\gamma^{}=E_\nu^{}/2$.\footnote{In the decay of $\nu_i^{}\to \nu_j^{}+a$, the energy of $\nu_i^{}$ is equally shared by two daughter particles $\nu_j$ and $a$ (for $m_i^{} \gg m_j^{}, m_a^{}$). In addition, after the ALP-photon conversion, the photon energy is equal to that of the ALP, i.e., $E_\gamma^{}=E_a^{}=E_\nu^{}/2$.}

\begin{figure}[t]
	\centering
	\includegraphics[scale=1]{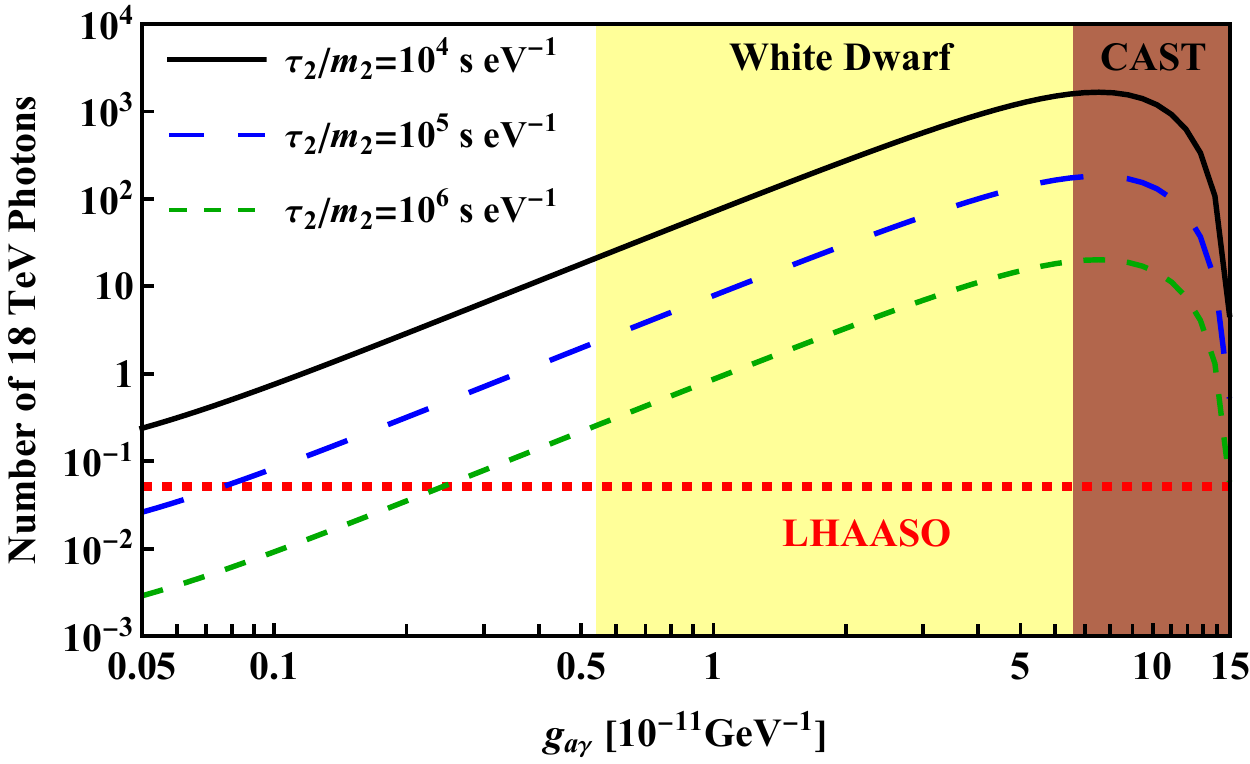}
	\caption{The expected number of $18~{\rm TeV}$ photons as the function of the ALP-photon coupling $g_{a\gamma}^{}$ for $\tau_2^{}/m_2^{} = 10^4~{\rm s}~{\rm eV}^{-1}$ (black solid curve), $10^5~{\rm s}~{\rm eV}^{-1}$ (blue long-dashed curve) and $10^6~{\rm s}~{\rm eV}^{-1}$ (green short-dashed curve). The Lorentz factor $\Gamma = 730$, the electron-to-proton energy ratio $f_e^{} = 0.01$, the ALP mass $m_a^{} = 10^{-10}~{\rm eV}$, and the neutrino lifetime $\tau_3^{} / m_3^{} = 10^7~{\rm s}~{\rm eV}^{-1}$ have been input. The exclusion region $g_{a\gamma}^{} \gtrsim 6.6 \times 10^{-11}~{\rm GeV}^{-1}$ ($95\%$ C.L.) from the CAST experiment~\cite{CAST:2017uph} and that $g_{a\gamma}^{} \gtrsim 5.4 \times 10^{-12}~{\rm GeV}^{-1}$ ($95\%$ C.L.) from magnetic white dwarfs~\cite{Dessert:2022yqq} are shaded in purple and yellow, respectively. The horizontal red dotted line corresponds to the expected event number of 0.0513, which is the lower limit for one event detected by LHAASO-KM2A~\cite{GCN32677} with 95\% C.L. based on the Poisson statistics~\cite{Gehrels:1986mj}.}
	\label{fig:g_a_gamma_18TeV_number}
\end{figure}
In Fig.~\ref{fig:g_a_gamma_18TeV_number}, we show the predicted number of $18~{\rm TeV}$ photons detected by LHAASO-KM2A in our scenario as the function of the ALP-photon coupling. In the numerical calculation, the Lorentz factor $\Gamma=730$, the electron-to-proton energy ratio $f_e^{} = 0.01$, the ALP mass $m_a^{}=10_{}^{-10}~{\rm eV}$ and the lifetime of the heaviest neutrino $\tau_3^{} / m_3^{} = 10^7~{\rm s}~{\rm eV}^{-1}$ have been fixed, while the $\nu_2^{}$ lifetime has been chosen as $\tau_2^{} / m_2^{} = 10^4~{\rm s}~{\rm eV}^{-1}$ (black solid curve), $\tau_2^{} / m_2^{} = 10^5~{\rm s}~{\rm eV}^{-1}$ (blue long-dashed curve), and $\tau_2^{} / m_2^{} = 10^6~{\rm s}~{\rm eV}^{-1}$ (green short-dashed curve). It is evident that the event number will be larger for shorter neutrino lifetimes and larger ALP-photon couplings. However, for large $g_{a\gamma}^{}$, the number of photons begins to oscillate due to the oscillatory behavior of the ALP-photon conversion probability $P_{a\gamma}^{}$, as indicated in the left panel of Fig.~\ref{fig:prob_g_ma}. The upper bound on the ALP-photon coupling $g_{a\gamma}^{} \lesssim 5.4 \times 10^{-12}~{\rm GeV}^{-1}$ at $95\%$ confidence level (C.L.) has been derived in Ref.~\cite{Dessert:2022yqq} through the polarization measurements of thermal radiation from magnetic white dwarfs, while the axion telescope CAST has put an upper bound $g_{a\gamma}^{} \lesssim 6.6 \times 10^{-11}~{\rm GeV}^{-1}$ at $95\%$ C.L. ~\cite{CAST:2017uph}. These exclusion regions are shaded in yellow and purple, respectively, in Fig.~\ref{fig:g_a_gamma_18TeV_number}, where the horizontal (red dotted) line represents the lower limit of expected event number for one observed event based on the Poisson probability distribution, which is 0.0513 at 95\% C.L.~\cite{Gehrels:1986mj}. 

\begin{figure}[t]
	\centering
	\includegraphics[scale=0.4]{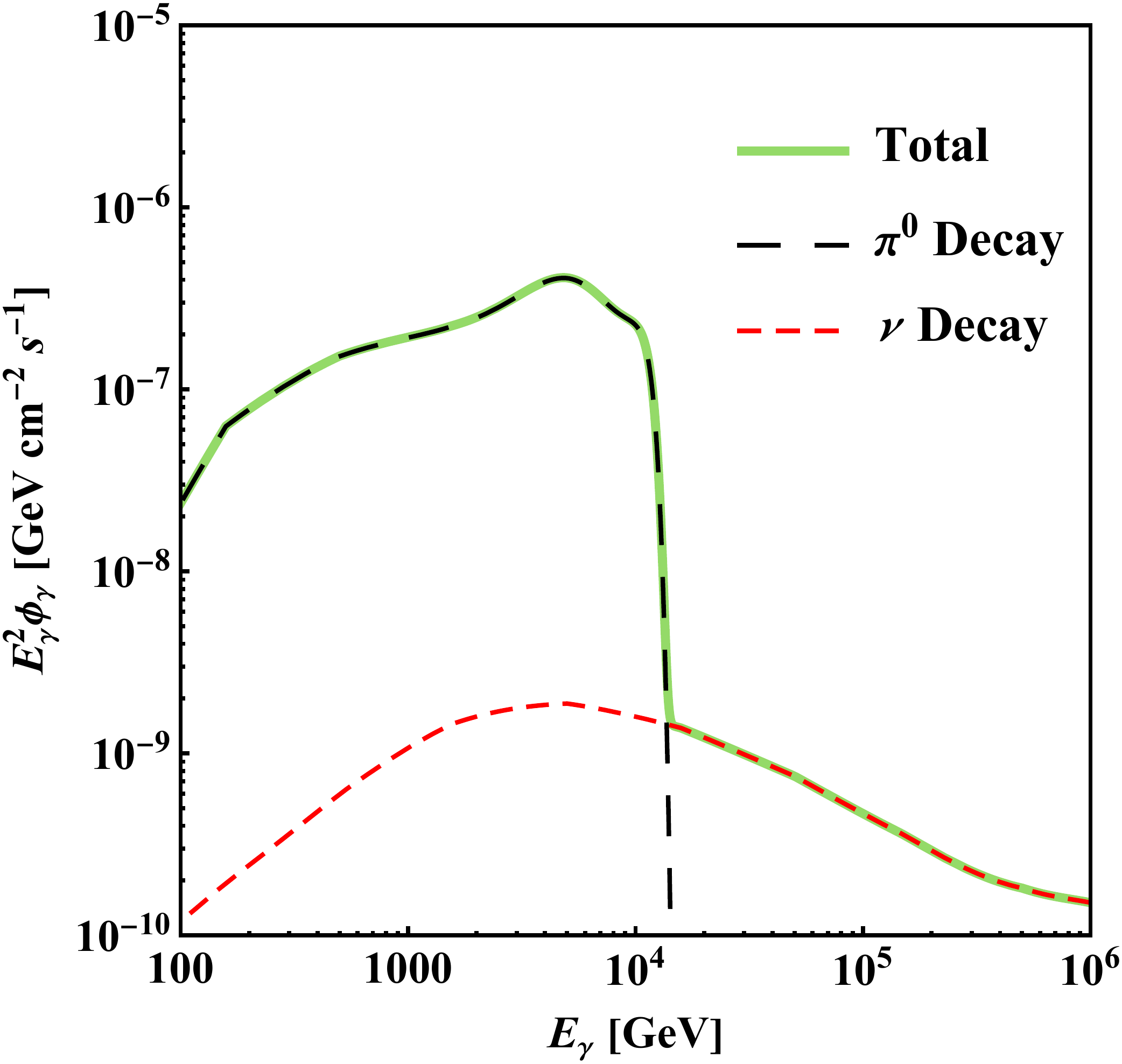}
	\caption{The fluxes of high-energy photons arriving at the detector from $\pi^0$ decays (black long-dashed curve) and from invisible neutrino decays (red short-dashed curve), while the total flux is represented by the green solid curve. The pion-induced photon flux is calculated by following Ref.~\cite{Hummer:2010vx} with the same input parameters as in Fig.~\ref{fig:Enu} and the EBL model is adopted from Ref.~\cite{Franceschini:2008tp}. Notice that the optical depth in the GRB internal shock is also taken into account~\cite{Murase:2022vqf}, which could significantly reduce the photon flux from $\pi^0$ decays. The neutrino-induced photon flux is calculated using the same input as above, namely, $\tau_2/m_2 = 10^4~{\rm s}~{\rm eV}_{}^{-1}$, $\tau_3/m_3 = 10^7~{\rm s}~{\rm eV}_{}^{-1}$, as well as the ALP mass $m_a = 10^{-10}~{\rm eV}$, and ALP-photon coupling $g_{a\gamma}^{}=10_{}^{-12}~{\rm GeV}_{}^{-1}$.}
	\label{fig:photon_pizero_neutrino}
\end{figure}

It has been pointed out in Ref.~\cite{Galanti:2022xok} that for most ALP interpretations of the LHAASO results, where photons are directly converted to ALPs at the GRB and then back to photons in the MW, a large value of $g_{a\gamma}^{}$ exceeding the upper bound from magnetized white dwarfs is needed. However, in our scenario, where the ALPs come from the invisible decays of GRB neutrinos, the desired values of $g_{a\gamma}^{}$ can be significantly smaller than the upper bound. In particular, for $\tau_2^{}/m_2^{} = 10_{}^{4}~{\rm s}~{\rm eV}^{-1}$ and $\tau_3^{}/m_3^{} = 10_{}^{7}~{\rm s}~{\rm eV}^{-1}$, we predict one event of $18~{\rm TeV}$ photons for $g_{a\gamma}^{} \approx 10_{}^{-12}~{\rm GeV}_{}^{-1}$, satisfying all the cosmological and astrophysical constraints on the ALP-photon coupling. On the other hand, given $g_{a\gamma}^{} \approx 10_{}^{-12}~{\rm GeV}_{}^{-1}$, if LHAASO does not observe more $18~{\rm TeV}$ photons, then a lower bound on $\tau_2^{}/m_2^{}$ can be inferred. This lower bound could be even stricter than that from cosmology.

\begin{figure}[t]
	\centering
	\includegraphics[scale=0.8]{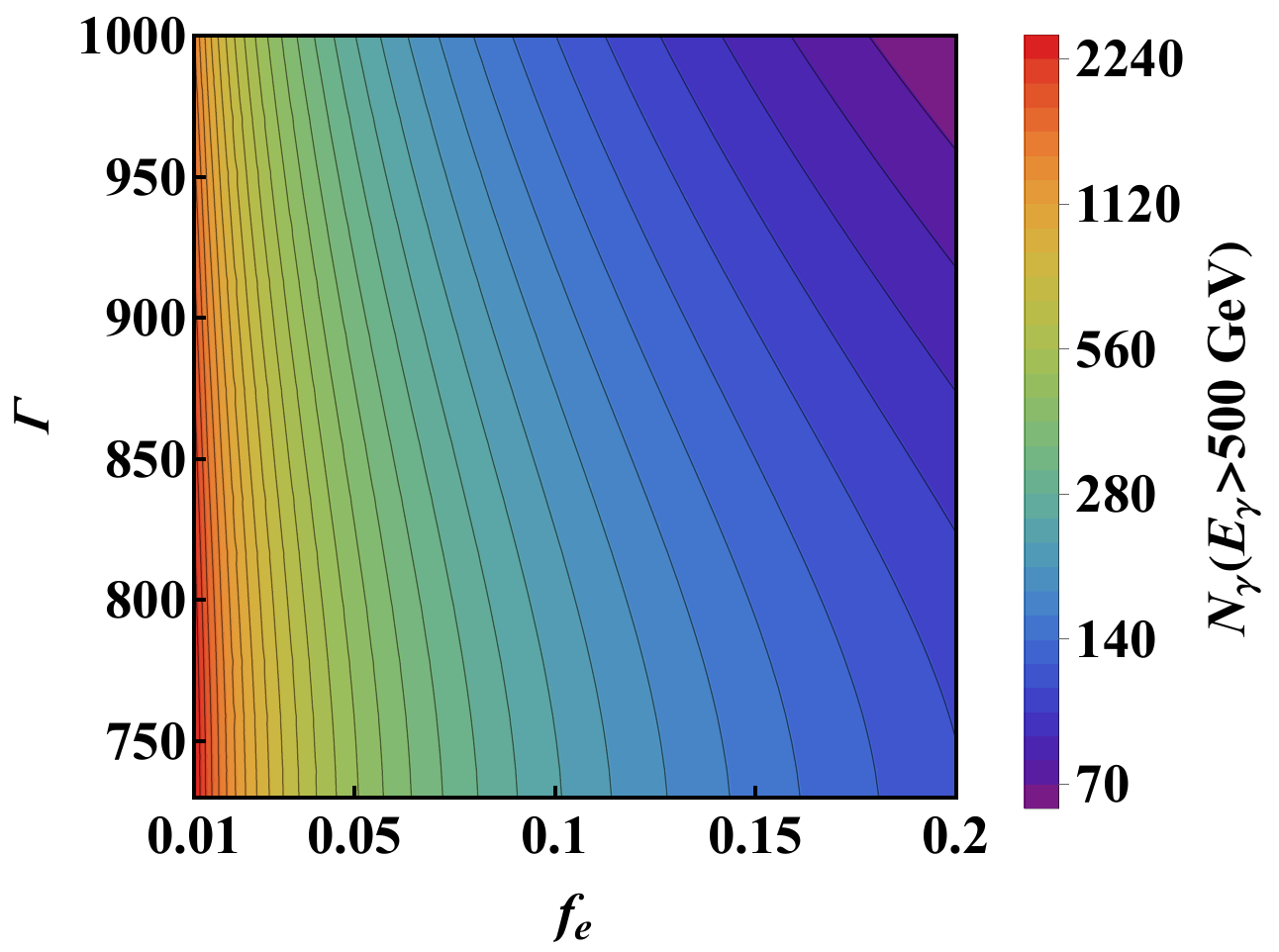}
	\caption{The total number of photons with energies higher than 500~GeV from the decays of both neutrinos and neutral pions, as the function of the Lorentz factor $\Gamma$ and the electron-to-proton energy ratio $f_e^{}$. The EBL model is adopted from Ref.~\cite{Franceschini:2008tp} and all the other input parameters are the same as those in Fig.~\ref{fig:photon_pizero_neutrino}.}
	\label{fig:totalnumbercontour}
\end{figure}
For the neutrino lifetime $\tau_2^{}/m_2^{} = 10_{}^{4}~{\rm s}~{\rm eV}^{-1}$, ALP-photon coupling $g_{a\gamma}^{}=10_{}^{-12}~{\rm GeV}_{}^{-1}$ and the same values of other parameters as in Fig.~\ref{fig:g_a_gamma_18TeV_number}, one can find out the total number of photons with energies above $500~{\rm GeV}$ by using Eq.~(\ref{eq:photon_number}), i.e., $N_{\gamma}^{\nu} \left(E_\gamma^{}>500~{\rm GeV}\right) \approx 60$, which is much smaller than the ${\cal O}(5000)$ photons reported by LHAASO~\cite{GCN32677}. However, it should be noted that high-energy photons are also produced from pion decays $\pi_{}^{0} \to 2\gamma$ in the GRB. In the right panel of Fig.~\ref{fig:Enu}, we have already presented the flux of gamma rays from neutral pion decays, both at the source and at the detector. Since the flux of high-energy photons will be attenuated in the EBL, we take the EBL model from Ref.~\cite{Franceschini:2008tp}. Such a model assesses the opacity information of the EBL in detail and has been well tested and widely applied in previous studies on GRBs, especially in the low-redshift region. In Fig.~\ref{fig:photon_pizero_neutrino}, the total flux of high-energy photons (green solid curve) from invisible neutrino decays (red short-dashed curve) and from neutral pion decays (black long-dashed curve) have been shown. As the optical depth $\tau^{}_{\rm op}(E_\gamma^{})$ becomes larger for higher-energy photons, the flux of pion-induced photons is exponentially suppressed for energies above $10~{\rm TeV}$. Taking into account the effective area of the LHAASO detectors~\cite{Ma:2022aau},  we can compute the predicted number of pion-induced photons with energies higher than $500~{\rm GeV}$, i.e.,
\begin{eqnarray}
N_{\gamma}^{\pi_{}^0}\left(E_\gamma^{}>500~{\rm GeV}\right)=\Delta t\int_{500~{\rm GeV}}^{100~{\rm TeV}} \phi_\gamma^{}\left(E_\gamma^{}\right) A_{\rm eff}^{}\left(E_\gamma
^{}\right){\rm e}_{}^{-\tau(E_\gamma^{})} \, {\rm d}E_{\gamma}^{}\approx 2200 \;,
\end{eqnarray}
where the upper limit of the integral is determined by the energy bound of the effective area of the LHAASO detectors. Meanwhile, the Lorentz factor and the electron-to-proton energy ratio have been fixed at $\Gamma=730$ and $f_e^{}=0.01$, corresponding to the maximum fluxes in the allowed range of $\Gamma$ and $f_e^{}$. Now consider the total number of photons coming from the decays of both neutrinos and neutral pions with different values of $\Gamma$ and $f_e^{}$. In Fig.~\ref{fig:totalnumbercontour} we show the prediction for the total number of photons with energies higher than 500~GeV to be detected by LHAASO, as the function of $\Gamma$ and $f_e^{}$. The total number is about 2300 at most in our model. Furthermore, it should be pointed out that high-energy gamma rays from the inverse Compton scattering of accelerated electrons off the background photons make an extra contribution. It is worthwhile to emphasize that the uncertainty in the EBL model is large, which may significantly affect the predicted number of pion-induced photons. For this reason, if a proper EBL model is implemented, the total number of VHE photons could be well consistent with the LHAASO observation. However, the prediction of ${\cal O}(10)~{\rm TeV}$ photons from neutrino decays in our scenario is independent of the EBL model. 

\section{Summary}
\label{sec:summary}
In this paper, we propose a novel and viable scenario of invisible neutrino decays $\nu^{}_i \to \nu^{}_j + a$ into ALPs and the ALP-photon conversion $a \to \gamma$ as the origin of ${\cal O}(10)~{\rm TeV}$ photons from GRB221009A. As GRBs serve as sources of high-energy cosmic rays, the accelerated protons will definitely interact with the ambient photons, producing both high-energy neutrinos and gamma rays via the photohadronic interactions. In this sense, both the original gamma rays and neutrinos from GRBs may contribute to the LHAASO observation of VHE photons from GRB221009A. 

Under the restrictive lower bounds on neutrino lifetimes from cosmology, we find that the conversion probability of neutrinos to ALPs can be sizable (e.g., $\gtrsim 30\%$ for TeV neutrinos). Furthermore, by taking into account the neutrino fluxes at the source and the ALP-photon conversion probability in the MW, we compute the expected number of VHE photons at LHAASO. Remarkably, an ${\cal O}(1)$ number of photons around $18~{\rm TeV}$ as observed by LHAASO can be predicted in our scenario, even with the ALP-photon coupling $g_{a\gamma}^{}\lesssim 10_{}^{-12}~{\rm GeV}^{-1}$ and the ALP mass $m_a^{}\sim 10^{-10}_{}~{\rm eV}$ that are compatible with all astrophysical constraints.

In comparison with other scenarios where high-energy photons are directly converted to ALPs at the GRB ~\cite{Galanti:2022pbg,Baktash:2022gnf,Lin:2022ocj,Troitsky:2022xso,Nakagawa:2022wwm,Zhang:2022zbm,Gonzalez:2022opy,Carenza:2022kjt,Galanti:2022xok}, ours is advantageous in two aspects. First, the interpretation of the ${\cal O}(10)~{\rm TeV}$ photons observed by LHAASO in our scenario depends neither on the magnetic field in the host galaxy nor on the concrete EBL models. Second, as pointed out in Ref.~\cite{Galanti:2022xok}, the viable parameter space of the ALP-photon coupling required to explain the TeV photons in most ALP models has some tensions with the upper bound from the magnetized white dwarfs, i.e., $g_{a\gamma}^{}\lesssim 5\times 10^{-12}_{}~{\rm GeV}_{}^{-1}$~\cite{Dessert:2022yqq}. Such a tension can be alleviated in our scenario.

In the near future, the energy spectrum of high-energy photons measured by LHAASO will be available. With the detailed energy spectrum, one can further explore all the allowed parameter space of the GRB models and particle-physics models. We look forward to such an interesting interplay among astronomy, astrophysics and particle physics.

\section*{Acknowledgements}

One of the authors (S.Z.) would like to thank Profs. Xiao-Jun Bi, Zhen Cao, Zhuo Li, Si-Ming Liu, Bo-Qiang Ma and Sai Wang for helpful discussions about the recent LHAASO observation. This work was supported in part by the National Natural Science Foundation of China under grant No. 11835013 and the Key Research Program of the Chinese Academy of Sciences under grant No. XDPB15.

\bibliographystyle{elsarticle-num}
\bibliography{ref}

\end{document}